\documentclass[11pt]{article}
\pdfoutput=1

\usepackage{times}
\usepackage{amsmath}
\usepackage{amssymb}
\usepackage{graphics}
\usepackage{graphicx}
\usepackage{verbatim}
\usepackage{xr}
\usepackage{rotating}
\usepackage{multirow}
\usepackage{float}
\usepackage[table]{xcolor}

\usepackage{cite}
\usepackage{setspace}
\usepackage[labelfont=bf,labelsep=period,justification=raggedright]{caption}

\makeatletter
\renewcommand{\@biblabel}[1]{\quad#1.}
\makeatother

\usepackage{hyperref}
\hypersetup{
  colorlinks,
  urlcolor=blue,
  linkcolor=black,
  citecolor=black
}

\floatstyle{boxed}
\newfloat{bx}{thp}{bx}
\floatname{bx}{Box}

\newcommand{\ws}[0]{W$\rightarrow$S}
\newcommand{\sw}[0]{S$\rightarrow$W}
\newcommand{\vect}[1]{\boldsymbol{\mathbf{#1}}}

\begin{document}

\title{\vspace{-6ex}A Model-Based Analysis of GC-Biased Gene Conversion \\ 
  in the Human and Chimpanzee Genomes} \author{John A. Capra$^{1,a,\dagger}$,
  Melissa J.\ Hubisz$^{2,\dagger}$, Dennis Kostka$^{3}$, Katherine
  S. Pollard$^{1,4,\ast}$, and Adam Siepel$^{2,\ast}$} \date{ }
\maketitle

\vspace{-6ex}
\begin{center}
\begin{footnotesize}
$^1$Gladstone Institutes, University of California, San Francisco, CA
94158, USA\\
$^2$Dept.\ of Biological Statistics and Computational Biology, Cornell
University, Ithaca, NY 14853, USA\\
$^3$ Depts. of Developmental Biology and Computational \& Systems Biology, University of Pittsburgh, Pittsburgh, PA 15201 USA\\
$^4$Institute for Human Genetics and Division of Biostatistics, University
of California, San Francisco, CA 94107, USA\\
$^a$\textit{Current Address:} Dept. of Biomedical Informatics and Center for Human Genetics Research, Vanderbilt University, \\ Nashville, TN 37232 USA\\
$^\dagger$These authors contributed equally to this work.\\
$^\ast$To whom correspondence should be addressed; E-mail:
acs4@cornell.edu; kpollard@gladstone.ucsf.edu.\\
\end{footnotesize}
\end{center}
\vspace{0.5ex}

\enlargethispage{\baselineskip}

\begin{center}
\begin{Large} {\bf Abstract} \end{Large}
\end{center}
GC-biased gene conversion (gBGC) is a recombination-associated process that
favors the fixation of G/C alleles over A/T alleles.  In mammals, gBGC is
hypothesized to contribute to variation in GC content, rapidly evolving
sequences, and the fixation of deleterious mutations, but its prevalence
and general functional consequences remain poorly understood.  gBGC is
difficult to incorporate into models of molecular evolution and so far has
primarily been studied using summary statistics from genomic comparisons.
Here, we introduce a new probabilistic model that captures the joint
effects of natural selection and gBGC on nucleotide substitution patterns,
while allowing for correlations along the genome in these effects.  We
implemented our model in a computer program, called phastBias, that can
accurately detect gBGC tracts $\sim$1~kilobase or longer in simulated
sequence alignments.  When applied to real primate genome sequences,
phastBias predicts gBGC tracts that cover roughly 0.3\% of the human and
chimpanzee genomes and account for 1.2\% of human-chimpanzee nucleotide
differences.  These tracts fall in clusters, particularly in subtelomeric
regions; they are enriched for recombination hotspots and fast-evolving
sequences; and they display an ongoing fixation preference for G and C
alleles.  We also find some evidence that they contribute to the fixation of
deleterious alleles, including an enrichment for disease-associated
polymorphisms.  These tracts provide a unique window into historical
recombination processes along the human and chimpanzee lineages; they supply
additional evidence of long-term conservation of megabase-scale
recombination rates accompanied by rapid turnover of hotspots.  Together,
these findings shed new light on the evolutionary, functional, and disease
implications of gBGC.  The phastBias program and our predicted tracts are
freely available.

\thispagestyle{empty}

\clearpage

%\com{Red text in brackets $=$ Tony's comments and questions.}

%\mtodo{Magenta text in brackets $=$ stuff that Tony thinks Melissa/Adam have done/are doing.}

%\mcom{Cyan text in brackets $=$ Melissa's comments and questions.}

%\acom{Green text in brackets $=$ Adam's comments.}

%\kcom{Orange text in brackets $=$ Katie's comments.}

%\clearpage

\section*{Introduction}

Gene conversion is the nonreciprocal exchange of genetic information from a
`donor' to an `acceptor' sequence, primarily resulting from the repair of
mismatched bases in heteroduplex recombination intermediates during meiosis
\cite{CHENETAL07B}.  In many cases, the process of resolving mismatches
between G/C (guanine or cytosine; denoted `strong' or `S') and A/T (adenine
and thymine; `weak' or `W') alleles appears to be biased in favor of S
alleles \cite{MARA03,CHENETAL07B,DUREGALT09}.  Such GC-biased gene
conversion (gBGC) elevates the fixation probabilities for S alleles
relative to W alleles at positions of W/S polymorphism, and if it acts in a
recurrent manner over a sufficiently long time, it can result in a
significant excess of \ws{} over \sw{} substitutions and a consequent
increase in equilibrium GC content.  It has been known since the 1980s both
that gene conversion occurs in various eukaryotes \cite{LAMB84}
%including several fungi, {\em Drosophila}, and maize \cite{LAMB84}---
and that mismatch repair can be significantly biased \cite{BROWJIRI88}.
%both
%between alleles of the same genomic locus and between homologous loci.
As complete genome sequences have become widely available,
evidence has accumulated that gBGC played an important role in genomic
evolution across many branches of the tree of life.  In particular, it is
likely that gBGC significantly influenced the genomic distribution of
GC content, the fixation of deleterious mutations, and rapidly evolving
sequences in many species
\cite{GALTDURE07,BERGETAL09,GALTETAL09,RATNETAL10,GLEM10,NECSETAL11,CAPRPOLL11,KOSTETAL12}.

Aside from limited experimental evidence of a GC-bias in meiosis, mostly
from yeast, much of what is known about gBGC comes from two indirect
sources of information: global patterns of variation within or between
species suggesting a fixation bias favoring S alleles 
\cite{DRESETAL07,DUREARND08,NECSETAL11,KATZETAL11,CAPRPOLL11} and the
existence of numerous loci exhibiting dense clusters of substitutions with
a pronounced \ws{} bias \cite{BERGETAL09,GALTETAL09,RATNETAL10,KOSTETAL12}.
Both types of evidence correlate strongly with recombination rates,
consistent with a cause from gBGC, although other 
recombination-associated factors might also contribute \cite{DUREARND08}.
However, these observations provide limited information about the
general prevalence, strength, and functional consequences of gBGC in humans
and other mammals.  Genome-wide patterns of variation are influenced by
diverse forces that act in a highly heterogeneous manner across the genome,
%---including mutation rates, direct selection, and selection from
%linked sites---
and it is difficult to measure the specific contribution of gBGC.  Clusters
of biased substitutions provide more direct evidence of a local
influence from gBGC.  So far, however, such clusters have been identified
by considering 
either genomic windows of fixed size or pre-identified genomic segments
(such as protein-coding exons or fast-evolving noncoding regions), which
has limited the regions that can be detected.  In
addition, many studies have focused on small numbers of clusters showing
extreme substitution rates and \ws{} biases.

For modelers of molecular evolution, gBGC is an anomaly---a process
separate and distinct from the fundamental processes of mutation,
recombination, drift, and selection that underlie most models, yet one
with the potential to profoundly influence patterns of variation within and
between species.  Like selection, gBGC acts in the window between the
emergence of genetic polymorphism due to mutation and its elimination due
to the fixation or loss of derived alleles. Unlike selection, however, gBGC
is
neutral with respect to fitness.  The influence of gBGC at individual
nucleotides can be modeled approximately by treating it as a selection-like
force that depends only on whether a new mutation is \ws{}, \sw{}, or neither
\cite{NAGY83,DUREARND08,KOSTETAL12}.  However, this approach ignores the
close association of gBGC with the notoriously difficult-to-model process
of recombination, which leads to a complex correlation structure along the
genome (i.e., gBGC ``tracts'' separated by regions of no gBGC).  Owing to
these difficulties, with only a few exceptions
\cite{RATNETAL10,KOSTETAL12}, gBGC has been ignored in phylogenetic or
population genetic models, and considered at most in post hoc
analyses (e.g., by examining identified genomic regions for an excess of
\ws{} substitutions).  These approaches are clearly limited in efficiency
and effectiveness, and there is a need for improved models of gBGC that can
be applied on a genome-wide scale.  There is also a need for high quality
annotations of gBGC-affected regions for use by 
investigators when interpreting comparative and population
genomic analyses.

Another reason to develop improved models of gBGC is that gBGC-induced
nucleotide substitutions provide a unique window into historical
recombination processes, by serving as a proxy for average
recombination rates along a lineage of interest.  By contrast, the other
main sources of information about recombination---sperm typing
\cite{JEFFETAL98}, genotypes for known pedigrees \cite{KONGETAL10}, and
patterns of linkage disequilibrium in present-day populations
\cite{MYERETAL05}---provide information about recombination that goes back
no farther than the coalescence time between individuals.  Pronounced
differences between the human and chimpanzee recombination maps suggest
that recombination rates in hominoids have changed rapidly
\cite{PTAKETAL05,WINKETAL05,AUTOETAL12}.  gBGC may provide useful
information about the recombination processes during the critical period
between the divergence of humans and chimpanzees (4--6 million years ago [Mya]) and the
coalescence time for human individuals ($\sim$1 Mya, on 
average).  (Archaic hominin genome sequences are of limited use for this
purpose, because they are still few in number and would extend
coalescence times only slightly.)

In this article, we address these issues by introducing a novel model-based
approach for the identification of gBGC tracts.  Our approach makes use of
statistical phylogenetic models that jointly consider gBGC and natural
selection \cite{KOSTETAL12}.  In addition, it approximates the
recombination-associated correlation structure of gBGC along the genome
using a hidden Markov model.  We have implemented this approach in a
computer program called phastBias, which is available as part of the
open-source {\em PH}ylogenetic {\em A}nalysis with {\em S}pace/{\em T}ime
models (PHAST) software package
(\url{http://compgen.bscb.cornell.edu/phast}) \cite{HUBIETAL11}.  Using
simulations, we show that phastBias can identify tracts of various lengths
from unannotated multiple alignments with good power.  We then analyze
genome-wide predictions of gBGC tracts in the human and chimpanzee genomes,
comparing them with recombination rates, patterns of polymorphism,
functional elements, fast-evolving sequences, and other genomic features.
This analysis sheds light on the prevalence and fitness consequences of
gBGC, and on recombination processes during the time since the
human/chimpanzee divergence.  
%In addition to our software, 
Our predictions of gBGC tracts are freely available as browser tracks
(\url{http://genome-mirror.bscb.cornell.edu}). We anticipate that these
tracks will 
be useful for avoiding false positives in scans for positive
selection, understanding the evolution of specific loci, and investigating
the broader evolutionary forces shaping the human genome. 

\section*{Results}

\subsection*{Probabilistic Model}

We model gBGC tracts using a phylogenetic hidden Markov model (phylo-HMM)
with four states, representing all combinations of gBGC or no gBGC in a
specified ``target'' genome (e.g., human or 
chimpanzee), and of evolutionary
conservation or no evolutionary conservation across the phylogeny
(Figure~\ref{fig:hmm}; 
Methods).  The phylo-HMM framework 
\cite{SIEPHAUS05} allows the distinct rates and patterns of nucleotide
substitution for each state to be described using a full statistical
phylogenetic model, and it captures the pronounced correlations along the
genomes in these patterns using a first-order Markov model.  Our phylo-HMM
can be thought of as a straightforward generalization of the two-state
model used by the phastCons program for prediction of evolutionarily
conserved elements \cite{SIEPETAL05} that additionally predicts gBGC tracts
in the target genome. 
We directly consider evolutionary
conservation together with gBGC because the dramatic reduction in
substitution rates in functional elements would otherwise be a confounding
factor in the identification of gBGC tracts.  The model allows conserved elements and gBGC tracts to
overlap or occur separately.  
The joint effects of gBGC and
selection are modeled by treating gBGC as a selection-like force that
specifically favors the fixation of G and C alleles, as in other recent
work.  In particular, the influence of selection is described using a 
population-scaled selection coefficient, $S=4N_es$, and the influence 
of gBGC is described using an analogous population-scaled 
GC-disparity parameter, $B=4N_eb$ (where $N_e$ is the effective population
size) \cite{KOSTETAL12} (see also \cite{NAGY83,DUREARND08}).  
The parameter $B$ measures the strength of gBGC, and values $B>0$ cause 
\ws{} substitution rates to increase and \sw{} substitution
rates to decrease. 
A key feature of our approach is that
it permits identification of gBGC tracts of any length based on
characteristic substitution patterns, independent of predefined windows or
genomic annotations.

Because the signal for gBGC in the data is typically quite weak, we make
several assumptions to reduce the complexity of the model.  Briefly, we
model negative selection as uniformly decreasing evolutionary rates on all
lineages, we ignore positive selection, and we assume that the disparity
parameter $B$ is the same for all gBGC tracts.  In addition, we
pre-estimate the parameters describing the neutral phylogeny and
evolutionary conserved elements using restricted models, we fix the
tract-length parameter $\alpha$ based on our prior expectation for tract
lengths, and we treat the parameter $B$ as a ``tuning'' parameter to be set
by trial and error (see summary of model parameters in Table
\ref{tab:hmmParameters}).  Our simulation study indicates that fairly high
accuracy in tract prediction is possible despite these simplifying
assumptions and approximations (see below and Methods for details).  We
have implemented our model in a program called phastBias in the PHAST
package \cite{HUBIETAL11}.  PhastBias makes use of existing features in
PHAST for alignment processing, phylogenetic modeling, efficient HMM-based
inference, and browser track generation.

\subsection*{Simulation Study}
While the absence of high-quality annotations of gBGC tracts makes it
difficult to assess prediction accuracy, we are able to gain some insight
into the performance of phastBias using simulated data.  To make our
simulated data as realistic as possible, we started with real genome-wide
alignments, and simulated new human sequences only, using our phylogenetic
model to define neutral and conserved sequences, and interspersed gBGC
tracts of fixed lengths (see Methods).  This strategy ensures that features
such as variation in mutation rates, changes in equilibrium GC content,
conserved elements, indels, alignment errors, and missing data are all
retained in 
the nonhuman sequences.  We used phastBias to predict human-specific tracts
based on these partially simulated alignments and compared our predictions
with the ``true'' tracts assumed during simulation.  We found that the
nucleotide-level false
positive rate was always very low in these experiments ($<4 \times 10^{-3}$
/ bp), so we measured the specificity of our predictions using the
nucleotide-level positive predictive value (PPV), defined as the fraction
of all bases predicted to be in gBGC tracts that truly belong in gBGC
tracts.  As a measure of power, we used the nucleotide-level true positive
rate (TPR), the fraction of bases in true gBGC tracts that were correctly
predicted as being in tracts.

First, we explored the performance of phastBias on simulated gBGC tracts of
various lengths, generated with several different values of the GC-disparity
parameter (denoted $B_{\text{sim}}$). Under our model, increasing
$B_{\text{sim}}$ produces tracts with more substitutions and greater GC
bias in their substitution patterns.  As expected, both our power to detect
gBGC and the specificity of our predictions increases with the lengths of
the true tracts and with $B_{\text{sim}}$ (Figure~\ref{fig:possim}). We
found that power and specificity were both quite good for tracts of
1,000--1,500 bases or longer, provided gBGC is reasonably strong
($B_{\text{sim}}\geq 5$).  
Current estimates of the lengths and GC-disparity of real gBGC tracts
\cite{WEBBETAL08,GALTETAL09} suggest that phastBias should have 
good power for many tracts (see Discussion).

Next, we examined the influence on prediction performance of our choice of
the tuning parameters for expected tract-length ($\alpha$) and gBGC
strength ($B$).  We found that the performance of the method was not highly
sensitive to the value of $\alpha$, so we decided to fix the expected tract
length at 1 kilobase (kb) (by setting $\alpha = 1/1000$) based on empirical evidence
indicating that mammalian gene conversion tracts are approximately this
size \cite{CHENETAL07B,WEBBETAL08}.  By contrast, the choice of $B$ had a
much stronger influence on the observed prediction performance. Power was
highest for small values of $B$, regardless of the value used to simulate
the tracts ($B_{\text{sim}}$) (Supplementary Figure
\ref{fig:morePosSims}).  However, this increase in power comes at only
a modest cost in PPV, which remains fairly high ($>$90\%) except when the
elements are both short and under weak gBGC (e.g., mean lengths of 100,
$B_{\text{sim}}=3$).  These results suggest that phastBias is inherently
somewhat conservative with its predictions, and that setting $B$ to a
relatively low value helps to improve sensitivity for tracts having a range
of true gBGC intensities, at minimal cost in specificity.

\subsection*{Predicted gBGC Tracts}

We applied phastBias to genome-wide alignments of the human, chimpanzee,
orangutan, and rhesus macaque genomes, and used it to predict tracts in the human
and chimpanzee genomes likely to have experienced gBGC since the divergence
of these two species 4--6 Mya (see Methods).  In separate runs, we selected either the
human or the chimpanzee genome as the ``target,'' and we set the tuning
parameter $B$ to values of 2, 3, 4, 5, and 10 (in increasing strength of
gBGC).  As expected from our simulation study, the number, lengths, and
genomic coverage of the predicted tracts depend fairly strongly on the
choice of $B$.  In particular, coverage decreases from more than 1\% to
0.07\% as $B$ is increased from 2 to 10 (Table~\ref{tab:hmmTractStats}).
Because the tracts predicted with high $B$ are largely found within those
predicted with lower $B$ (Supplementary
Table~S2 %~\ref{tab:tract_coverage}), 
and because a value of $B=3$ appears
to result in good power while controlling false positives (see above), we
will focus on the tracts predicted with $B=3$ for the remainder of the
article.  The absolute sensitivity of these predictions of course depends
on unknown properties of true gBGC tracts, but our simulation experiments
indicate that power is fairly good, at least for the subset of tracts
$\geq$1 kb in length with a reasonably pronounced GC-disparity (Figure
\ref{fig:possim}).

With $B=3$, the predictions for the human genome include 9,439 gBGC tracts
covering 0.33\% of the genome (Table~\ref{tab:hmmTractStats}).  These
predicted tracts average 1,018 bp in length (median 788 bp), consistent
with our choice of $\alpha = 1/1000$, but they display a fairly broad
length distribution (Supplementary Figure~\ref{fig:length_dist}), indicating
that our choice of tuning parameters is not overly restrictive.  Most
predicted tracts contain exclusively or predominantly \ws{} substitutions
(Supplementary Figure~\ref{fig:bias_dist}).  The statistics for the
chimpanzee genome are similar, but in this case there are somewhat fewer
tracts (8,677), their lengths are reduced (mean = 842 bp, median = 663 bp),
and genomic coverage is about 25\% lower (at 0.25\%).  The reduced coverage
of the chimpanzee genome holds even if we consider only tracts that
completely fall within regions of high-quality, syntenic alignment between
the two genome assemblies.  These differences between the human and
chimpanzee predictions could reflect differences between species in the
degree to which recombination events are concentrated in recombination
hotspots \cite{AUTOETAL12} (see Discussion).

The human predictions are distributed across the human genome but show a
clear tendency to cluster near the ends of chromosomes
(Figure~\ref{fig:human_v_chimp}), consistent with previous
findings~\cite{DRESETAL07,ROMIETAL10,CAPRPOLL11}.  The median distance from
the nearest telomere is only about one third that observed for a set of
GC-content-matched control regions (9.6 megabases (Mb) vs.\ an average of 30.4 Mb over
1000 replicates, $p<0.001$; see Methods and Supplementary Figure~\ref{fig:qqplots}).  Similarly, the median distance between tracts is
less than one third that for the controls, even after merging tracts $<$1
kb apart to account for possible biases from the HMM-based prediction method (24.3
kb vs.\ an average of 86.0 kb, $p<0.001$). In addition to the clusters
of tracts near present-day telomeres, there is an obvious cluster near the
centromere of chromosome 2, reflecting the telomeres of two ancestral
chromosomes that fused at this site sometime after the human/chimpanzee
divergence~\cite{IJDOETAL91,DRESETAL07}.  The genomic distribution of the
predicted tracts in chimpanzee is generally similar to that in human
(Figure~\ref{fig:human_v_chimp}, Supplementary Material).

Together, the human and chimpanzee tracts account for about 1.2\% of all
human/chimpanzee nucleotide differences apparent in our genome-wide
alignments (435,729 differences).  About half (214,195) of the nucleotide
differences within the tracts can be confidently explained by
\ws{} substitutions on either the human or chimpanzee lineage, of
which slightly more than half (115,699) fall on the human lineage.  Thus,
even with our limitations in power, our predictions suggest a non-negligible
influence of gBGC on overall levels of human/chimpanzee nucleotide
divergence.

\subsection*{Recombination Rates} 
The predicted human gBGC tracts are substantially enriched for
recombination hotspots from the HapMap project \cite{HAPMCONS07}: 1,228
(13\%) overlap a hotspot, compared with an average of 796 for the control
regions ($p < 0.001$).  In addition, the average recombination rate
\cite{1KGCONS10} within these tracts is more than twice the rate in the
control regions (3.85 centimorgans per megabase (cM/Mb) vs. 1.61 cM/Mb, $p
< 0.001$; Table~\ref{tab:human_chimp_recomb}).  A parallel analysis of the
chimpanzee gBGC tracts based on the genome-wide recombination rate map from
the PanMap Project~\cite{AUTOETAL12} showed, similarly, that recombination
rates in predicted gBGC tracts were more than twice as high as in control
regions (Table~\ref{tab:human_chimp_recomb}).  Pedigree-based human
recombination maps~\cite{KONGETAL10} produced similar results (data not
shown).

At fine scales, the human and chimpanzee tracts show a modest, but
significant, degree of overlap (Figure~\ref{fig:human_v_chimp}): 605
(6.4\%) of the human tracts directly overlap a chimpanzee tract, compared
with an average of 86 for the control regions ($p<0.001$).  Shared
recombination hotspots account for only a small minority ($<$1\%) of the
overlapping tracts.  However, the correlation in tract locations between species is much
stronger at broader scales.  For example, if the fractions of nucleotides
in gBGC tracts are compared in orthologous genomic blocks of various sizes,
the human/chimpanzee Pearson's correlation increases from $r=0.25$ for 10
kb blocks to $r=0.57$ for 100 kb blocks, and to $r=0.80$ for 1 Mb blocks (Supplementary Figure~\ref{fig:hc_tract_correspondence}).
These observations mirror those for human and chimpanzee recombination
rates, which correlate well at scales of 1 Mb or larger but much more
poorly at finer scales \cite{PTAKETAL05,WINKETAL05,AUTOETAL12}.

To gain further insight into the conservation of the gBGC tracts, we mapped
the human gBGC tracts to orthologous locations in the chimpanzee genome,
and the chimpanzee tracts to orthologous locations in the human genome.  We
then compared the recombination rates in these ``ortho-tracts'' with those
in control regions, as with the tracts predicted for each species.  Unlike
recombination hotspots \cite{AUTOETAL12}, the predicted gBGC tracts show
significantly elevated recombination rates at orthologous positions in the
other species (Table~\ref{tab:human_chimp_recomb}).  However, these
recombination rates are not nearly as elevated as those in the genome in
which the tracts were predicted.  Together, these observations indicate that,
while the human and chimpanzee gBGC tracts are only weakly correlated, they
nevertheless exhibit residual correlation beyond what is observed
with present-day recombination rates, probably because
they reflect average recombination rates over millions of years (see
Discussion). 

In both human and chimpanzee, the predicted tracts show a weak positive
correlation with GC-content on a megabase scale.  This correlation is
somewhat stronger for human (Pearson's correlation for 1 Mb blocks:
$r=0.12$) than for chimpanzee ($r=0.09$), mirroring observations of a
stronger correlation of recombination rate with GC-content in human than in
chimpanzee \cite{AUTOETAL12}.

\subsection*{Genomic Annotations}

To shed light on the functional implications of gBGC, we examined the degree of overlap of the predicted human gBGC tracts with
various sets of genomic annotations (listed in Methods).  
In comparison with the control regions, 
we found that the human gBGC tracts were significantly depleted for overlap
with known protein-coding exons, core promoters (1 kb upstream of annotated
transcription start sites), miscellaneous RNAs, LINEs and SINEs, while they
were significantly enriched for overlap with introns, lincRNAs, and a
collection of ChIP-seq-supported transcription factor binding sites
(Supplementary Figure~\ref{fig:enriched_features}).  However, all of these enrichments and
depletions were modest in magnitude, with fold-changes of about 0.8--1.3.
Overall, the gBGC tracts appear to be fairly representative of sequences of
the same GC content.
It is
possible that the depletion for gBGC tracts in protein-coding exons and
promoters could result in part from strong purifying selection 
counteracting GC-biased fixation.

\subsection*{GC-Bias in Derived Alleles} 

To distinguish between fixation- and mutation-related biases, we compared
the derived allele frequencies at polymorphic \ws{} and \sw{} sites in the
predicted tracts and control regions.  To control for the possibility of an
ascertainment bias from polymorphic sites at which the derived allele is
present in the human reference genome, we performed this analysis twice:
once with the original gBGC tracts, and once with predictions based on
alignments in which polymorphic sites in the human genome had been masked
with `N's.

\begin{figure}[h]
\begin{center}
\includegraphics[width=6.83in]{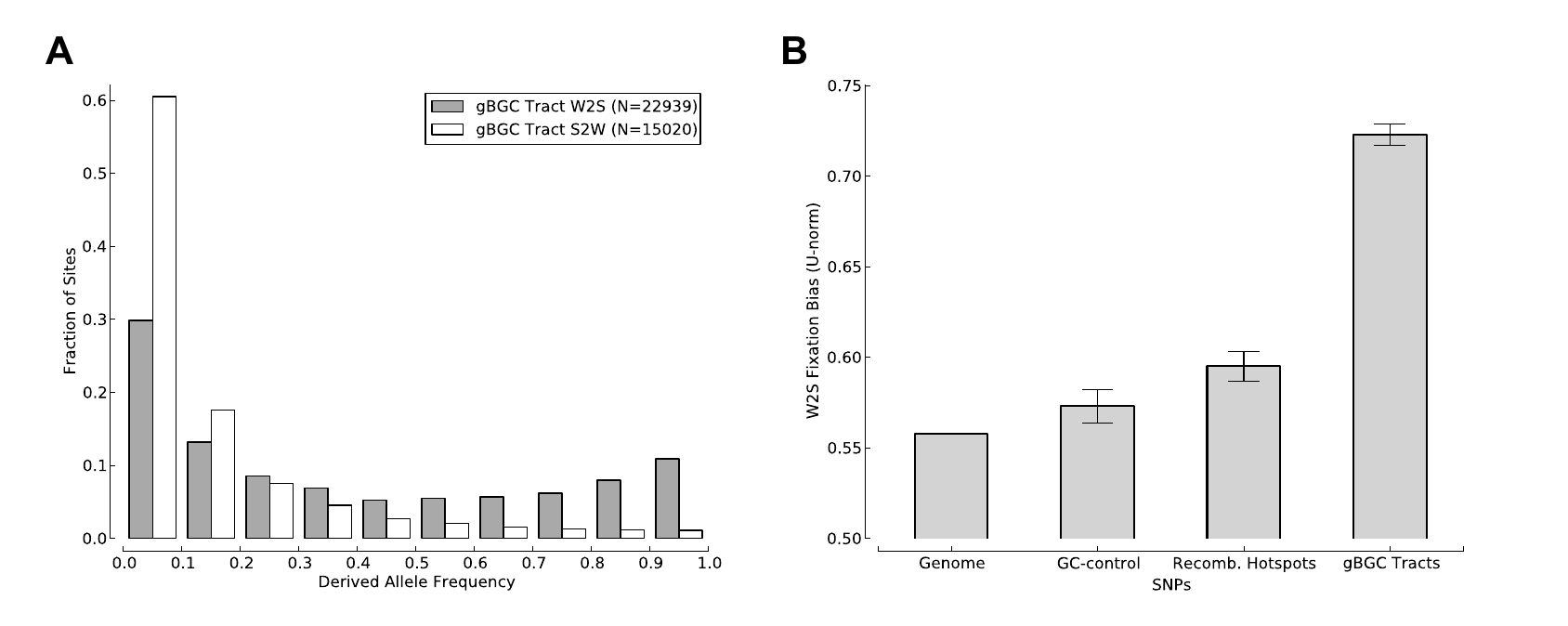}
\caption{\textbf{Human polymorphism data indicates an ongoing preference
    for the fixation of G and C alleles in the predicted gBGC tracts.}
  \textbf{(A)} \ws{} changes in gBGC tracts have significantly higher
  derived allele frequencies than \sw{} changes in tracts.  This plot is
  based on data for the YRI population from the 1000 Genomes
  Project~\cite{1KGCONS10}.  Results for other populations were similar
  (data not shown).  \textbf{(B)} The $U$-norm, a measure of the degree of
  \ws{} bias \cite{KATZETAL11}, is significantly higher in gBGC tracts than
  in the entire genome or in GC-matched control regions (see Methods).
  Recombination hotspots also show somewhat elevated values but much less
  elevated than the predicted tracts.  The error bars indicate 95\%
  confidence intervals.}
\label{fig:gbgc_daf_spectra}
\end{center}
\end{figure}

Based on data from the 1000 Genomes Project~\cite{1KGCONS10} (YRI
population), the predicted gBGC tracts displayed significantly elevated
derived allele frequencies at sites of inferred \ws{} mutations compared
with sites of inferred \sw{} mutations (\ws{} DAF skew of 0.723 $\pm$
0.006; Figure~\ref{fig:gbgc_daf_spectra}A).  This skew in DAFs was
significantly greater than that observed genome-wide (0.558 $\pm$ 0.001)
or in recombination hotspots (0.595 $\pm$ 0.008;
Figure~\ref{fig:gbgc_daf_spectra}B), and it was larger than observed in any
of the 1000 control region replicates (0.573 $\pm$ 0.009).  The tracts are
also far more biased than any of the regions considered by Katzman et
al.~\cite{KATZETAL11}, which were identified using sliding windows of fixed size and likely contained a mixture of gBGC tracts and non-tracts. 
Results were similar for the CEU (0.703 $\pm$ 0.007)
and CHB-JPT populations (0.678 $\pm$ 0.008).  These results held for the
tracts based on the polymorphism-masked alignments, although the magnitude
of the skew was somewhat reduced in this case (0.685 $\pm$ 0.007 for YRI;
Supplementary Figure~\ref{fig:ongoing_bias_nopoly}).  Together, these
results strongly indicate an on-going preference for the fixation of G and C
alleles in the predicted gBGC tracts.

\subsection*{Fixation of Deleterious Alleles}

Theoretical modeling has shown that gBGC, in principle, can overcome
negative selection and result in the fixation of weakly deleterious
alleles~\cite{GALTETAL09, DUREGALT09, GLEM10}.  However, there is currently
little direct empirical evidence of a contribution of gBGC to fixed or
segregating deleterious alleles~\cite{NECSETAL11}.  Our genome-wide tract predictions enabled
us to investigate the link between gBGC and deleterious alleles in two
ways: by testing for enrichments for disease-associated genomic regions in
gBGC tracts, and by examining evidence of purifying selection in
orthologous regions of other mammalian genomes.

First, we examined the relationship between the gBGC tracts and four sets
of putatively disease-assoc\-iated genomic regions: 10,711 polymorphic sites
from dbSNP annotated as ``pathogenic'' or ``probable path\-ogenic''
\cite{SHERETAL01}; 43,952 polymorphic sites from the Human Gene Mutation
Database (HGMD)~\cite{STENETAL09} (see also~\cite{NECSETAL11}); 11,444
genomic regions from the Genetic Association Database
(GAD)~\cite{ZHANETAL10}; and 6,435,165 polymorphic sites with evidence of
functional importance (classes 1--5) in RegulomeDB ~\cite{BOYLETAL12}.  For
the dbSNP pathogenic and HGMD comparisons, we considered sets of control
regions that overlapped the same number of exonic SNPs as the gBGC tracts.
This control is designed to avoid misleading findings of significance that
simply reflect the GC content, exon coverage, and/or rates of polymorphism
in the gBGC tracts, since these disease-associated region sets are mostly
found in coding regions. Similarly, we used control regions matched to SNPs
considered by RegulomeDB, since it only includes non-coding SNPs (Methods).
We found that the gBGC tracts overlapped significantly more putatively
disease-related SNPs from the dbSNP, HGMD, and RegulomeDB collections, and
significantly more of the GAD regions, than did the matched control regions
(Table~\ref{tab:disease_enrich}; $p < 0.05$ for each).

Second, we looked for evidence of purifying selection in chimpanzees and
other species at the locations of \ws{} substitutions within the predicted
human tracts.  If a substantial number of these
mutations were driven to fixation by gBGC despite negative selection
against them, we would expect to observe an excess of evolutionary
conservation, a 
deficiency of polymorphisms, and/or a skew toward low-frequency derived
alleles at orthologous locations in other species.  
However, we found that the bulk distributions of phyloP conservation scores
\cite{POLLETAL09}, computed for eutherian mammals but excluding human and
chimpanzee (Methods), were nearly identical for the tracts and the
GC-matched controls (Supplementary Figure~\ref{fig:phyloP}).  In
addition, we compared chimpanzee polymorphisms \cite{AUTOETAL12} in regions
orthologous to our human tracts and control regions, and found no deficiency of
polymorphisms (Supplementary Figure~\ref{fig:chimpPolyCount}) and 
no excess of low-frequency derived alleles (Supplementary Figure~\ref{fig:chimpPolyFreq}) within the tracts.  Indeed, the regions
orthologous to the human tracts displayed an excess of chimp polymorphisms, perhaps
reflecting increased power for gBGC detection in regions of elevated
mutation rates.  On the other hand, we did observe a significant enrichment
for overlap with evolutionarily conserved elements identified using
phastCons \cite{SIEPETAL05} at locations of \ws{} substitutions within the
predicted tracts (Supplementary Figure~\ref{fig:phastConsEnrich}).
Thus, we found mixed evidence linking gBGC with fixation of
deleterious alleles.

\subsection*{Overlap with Fast-Evolving Sequences}

Many fast-evolving regions of the human genome display an excess of \ws{}
substitutions, leading to the suggestion that gBGC may play a role in their
evolution~\cite{POLLETAL06B,GALTDURE07,KATZETAL10,KOSTETAL12,BERGETAL09,RATNETAL10}.
Supporting this hypothesis, our predicted gBGC tracts overlap 13 of the 202
(6.4\%) HARs identified by Pollard et al.\ \cite{POLLETAL06B}, more than
observed for any of the 1000 GC-control region replicates ($p < 0.001$).
Notably, the HARs overlapped by gBGC tracts included HAR1, HAR2, and HAR3,
the three fastest evolving sequences in this set.  We also examined an
expanded set of 721 HARs~\cite{LINDETAL12} and found that gBGC tracts
overlapped 75 of them (10\%; $p < 0.001$; see example in
Figure~\ref{fig:browser_example}).  Next, we compared the gBGC tracts with
the 10 protein-coding genes recently identified as showing signatures of
positive selection on the human branch based on a likelihood ratio test
\cite{KOSIETAL08}.  One of these genes is overlapped by a gBGC tract,
significantly more than expected based on exon-aware controls ($p=0.009$).
The overlapped gene, {\em ADCYAP1}, was also highlighted by another group
\cite{RATNETAL10} as showing strong evidence of an influence from gBGC.
We repeated our analysis with 157 genes identified in another recent study as
showing signatures of human-specific positive selection \cite{GEORETAL11},
and found that the gBGC tracts overlapped 11 (7\%) of these genes, somewhat
more than average for the exon-aware control replicates (7.4, $p =
0.077$).  Considering our limitations in power (see Discussion), these
results indicate the gBGC 
has contributed to a substantial fraction of fast-evolving sequences in the
human genome.

\subsection*{Genome Browser Track}

\begin{figure}[h]
\begin{center}
\includegraphics[width=6.83in]{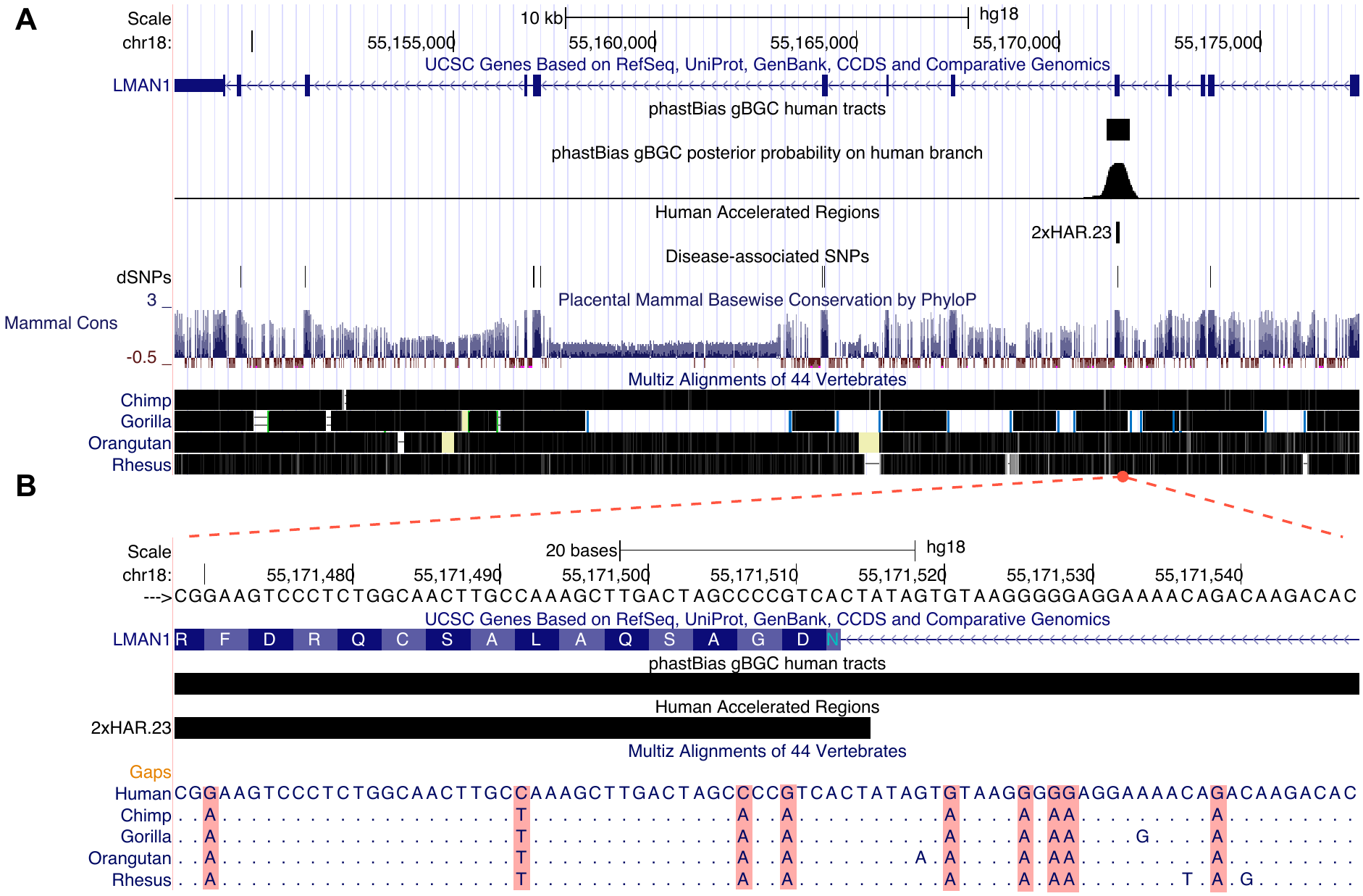}
\caption{\textbf{Illustration of genome browser track.}  \textbf{(A)} UCSC
  Genome Browser screen shot focused on the {\em LMAN1} gene (hg18.chr18:55,148,088-55,177,461). This region contains a predicted gBGC tract (black
  bar, second track from top); the ``wiggle'' track below shows the posterior probability of gBGC at each site computed by phastBias. The gBGC tract overlaps an exon of the gene
  (blue bar at top; adjacent chevrons indicate introns), a human
  accelerated region (2xHAR.23; short black bar), and a known missense
  variant from dbSNP (rs146465318; black tick mark).  The phyloP-based
  conservation track (``Mammal Cons'') shows that phastBias can predict
  tracts that span both conserved and nonconserved regions.  The phastBias track is
  available at \url{http://genome-mirror.bscb.cornell.edu} (hg18
  assembly). Notably, this region has an elevated recombination rate (2.5
  cM/Mb; not shown).  \textbf{(B)} The multiple sequence alignment for a
  portion of the gBGC tract (hg18.chr18:55,171,469-55,171,548) illustrates the characteristic signature of
  gBGC. This interval has nine human-specific \ws{} substitutions over 80
  nucleotides, four of which fall within the exon.  Positions in other
  species that match the human sequence are indicated with a period. }
\label{fig:browser_example}
\end{center}
\end{figure}

Our predicted tracts for human and chimpanzee 
are available as a UCSC Genome Browser track at
\url{http://genome-mirror.bscb.cornell.edu} (Figure~\ref{fig:browser_example}).  This track displays both our discrete
predictions of gBGC tracts and a continuous-valued plot indicating the
posterior probability that each position is influenced by gBGC.  Using
this track it is possible to browse the predicted tracts in their full
genomic context, perform queries intersecting them with other browser
tracks, and 
download them for further analysis.  We expect this track to
be particularly useful for other investigators who wish to exclude
gBGC-influenced regions of the genome from other molecular evolutionary
analyses, such as the identification of genes under positive selection. The tracts themselves will also be directly useful for studying the evolution of recombination rates and their relationship to substitution rates and patterns.

\section*{Discussion}

This paper describes an analysis of predicted gBGC tracts in the human and
chimpanzee genomes, based on a new computational method called phastBias.
PhastBias makes use of a hidden Markov model and statistical phylogenetic
models that consider the influence of both natural selection and gBGC on
substitution rates and patterns.  Unlike previous methods for identifying
signatures of gBGC, it does not depend on a sliding window or previous
genomic annotations, but instead can flexibly identify tracts of various
sizes directly from genome-scale multiple alignments.  The method appears
to have good power for tracts of about 1 kilobase or longer, provided gBGC
has acted with a reasonably high average intensity along the lineage of
interest.  Our predictions in the human and chimpanzee genomes cover about
0.3\% of each genome, and explain about 1.2\% of human/chimpanzee single
nucleotide differences.  Consistent with a cause from gBGC, the predicted
tracts are correlated with recombination rates, tend to fall in
subtelomeric regions, and exhibit an ongoing fixation bias for G and C
alleles.  In addition, they tend to overlap previously identified
fast-evolving coding and non-coding regions, suggesting
that gBGC has contributed significantly to the evolution of these
sequences.  Overall, our analyses indicate that gBGC has been an important
force in the evolution of human and chimpanzees since their divergence 4--6
million years ago.

\begin{figure}[h]
\begin{center}
\includegraphics[width=6.83in]{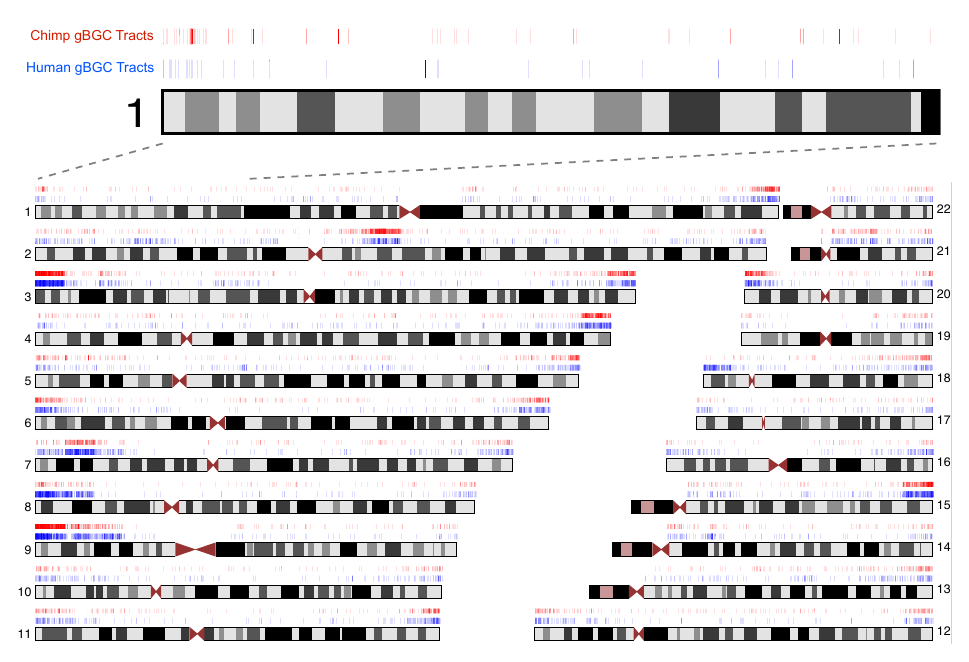}
\caption{\textbf{Genomic distribution of predicted human and chimpanzee
    gBGC tracts.}  Both human (blue) and chimpanzee (red) gBGC tracts are
  found throughout the genome, but tend to cluster and fall near
  telomeres. Chimpanzee gBGC tracts are displayed at the 
  corresponding aligned positions in the human genome. The dense cluster of
  gBGC tracts near the centromere of chromosome 2 is the site of the fusion
  of two ancestral chromosomes on the human lineage. This region is
  telomeric in chimpanzee and was telomeric for much of human evolution.  As
  illustrated by the magnified section of chromosome 1, human and
  chimpanzee tracts often occur in similar regions, but rarely overlap. }
\label{fig:human_v_chimp}
\end{center}
\end{figure}

Many attributes of the predicted gBGC tracts are consistent with
recombination as the driving force of gBGC.  Nevertheless, the tract
locations are only partially correlated with recombination rates in human
and chimpanzee.  Moreover, while the tracts are enriched for recombination
hotspots in both species, there are thousands of hotspots that do not
overlap a gBGC tract, and the majority of tracts do not overlap a hotspot.
These differences can be explained by several factors.  First, the hotspots
we have analyzed reflect recombination patterns in modern human
populations, while the gBGC tracts reflect average patterns since the
divergence of humans and chimpanzees. Many 
current hotspots presumably have not had sufficient time to produce a detectable
signature of biased substitution, while many extinct hotspots 
contributed to gBGC for long periods of time in the past.  The gBGC imprint
of such historical recombination hotspots can be observed most dramatically
near the fusion of chromosome 2 (Figure~\ref{fig:human_v_chimp}).  Second,
models of gBGC suggest that it can occur in conjunction with both crossover
and noncrossover recombination events, but current recombination maps
reflect crossover events only \cite{DUREGALT09}.  An imperfect correlation
of these types of events, together with statistical noise in current
estimates of crossover rates, likely accounts for some of the absence of
correlation between recombination rates and gBGC tracts.  Finally, biased
substitution rates are influenced by factors other than recombination, such
as mutation rates and natural selection.  For example, strong purifying
selection at or near a hotspot could eliminate the signature of gBGC.

The locations of the human and chimpanzee tracts are strongly correlated on
megabase scales, but, like recombination rates, they differ significantly
on fine scales, and few human and chimpanzee tracts directly overlap one
another (Figure~\ref{fig:human_v_chimp}; Supplementary Figure~\ref{fig:hc_tract_correspondence}).  Nevertheless, even at fine scales, the gBGC tracts for the two
species agree better than recombination hotspots, which are essentially
uncorrelated \cite{AUTOETAL12}.  This observation probably stems from the
fact that gBGC tracts reflect time-averaged recombination rates, and
historical recombination rates were presumably better correlated than those
in present-day humans and chimpanzees.  In general, the predicted gBGC
tracts provide a valuable window into historical recombination processes,
but this window is ``blurred'' by time-averaging over millions of years.
Nevertheless, together with other sources of information about historical
recombination processes---such as new methods based on patterns of
incomplete lineage sorting (K.\ Munch, T.\ Mailund, J.Y.\ Dutheil, and
M.H.\ Schierup, in prep.)---predictions of gBGC tracts may help to fill in
our picture of the evolution of recombination rates in hominoids.

Despite the overall similarity of the human and chimpanzee predictions, the
coverage of the predicted tracts is 
about 25\% lower in the chimpanzee genome.  A possible cause of this
difference is the greater concentration of recombination events in
hotspots in humans \cite{AUTOETAL12}.  This
difference could lead to a stronger population-level signal for gBGC in humans,
allowing for more predictions and longer predicted tract lengths.  Because
this concentration of recombination events is more pronounced for European
than for African populations, it may be of interest to predict separate
sets of tracts for individuals of European and African descent and see
whether an excess of gBGC tracts are observed in Europeans.  

The difference between humans and chimpanzees in the concentration of
recombination events may derive from differences in the activity of the
hotspot-specifying protein PRDM9, which shows substantially greater allelic
diversity in chimpanzees than in humans \cite{AUTOETAL12}.  
Consistent with this hypothesis,
there is a much weaker signal for sequence motifs potentially involved in
PRDM9 binding at chimpanzee hotspots than at human hotspots.
In an attempt to shed light on the ancestral binding
preferences of PRDM9, we applied motif discovery methods to the predicted
gBGC tracts in the human and chimpanzee genomes.  However, in both species
this analysis turned up only a few motifs, none of which resembled the
well-defined motifs reported for the human genome
\cite{MYERETAL08,AUTOETAL12}.  It may be that the ancestral recombination
hotspots in both species are more like the ones in present-day chimpanzees
than those in present-day humans, with poorly defined sequence motifs.
Alternatively, the absence of well-defined motifs may simply reflect
rapidly evolving PRDM9 binding preferences and the time-averaged nature of
the gBGC tracts, which together could eliminate a clear signal for motif
discovery.

\begin{figure}[h]
\begin{center}
\includegraphics[width=4in]{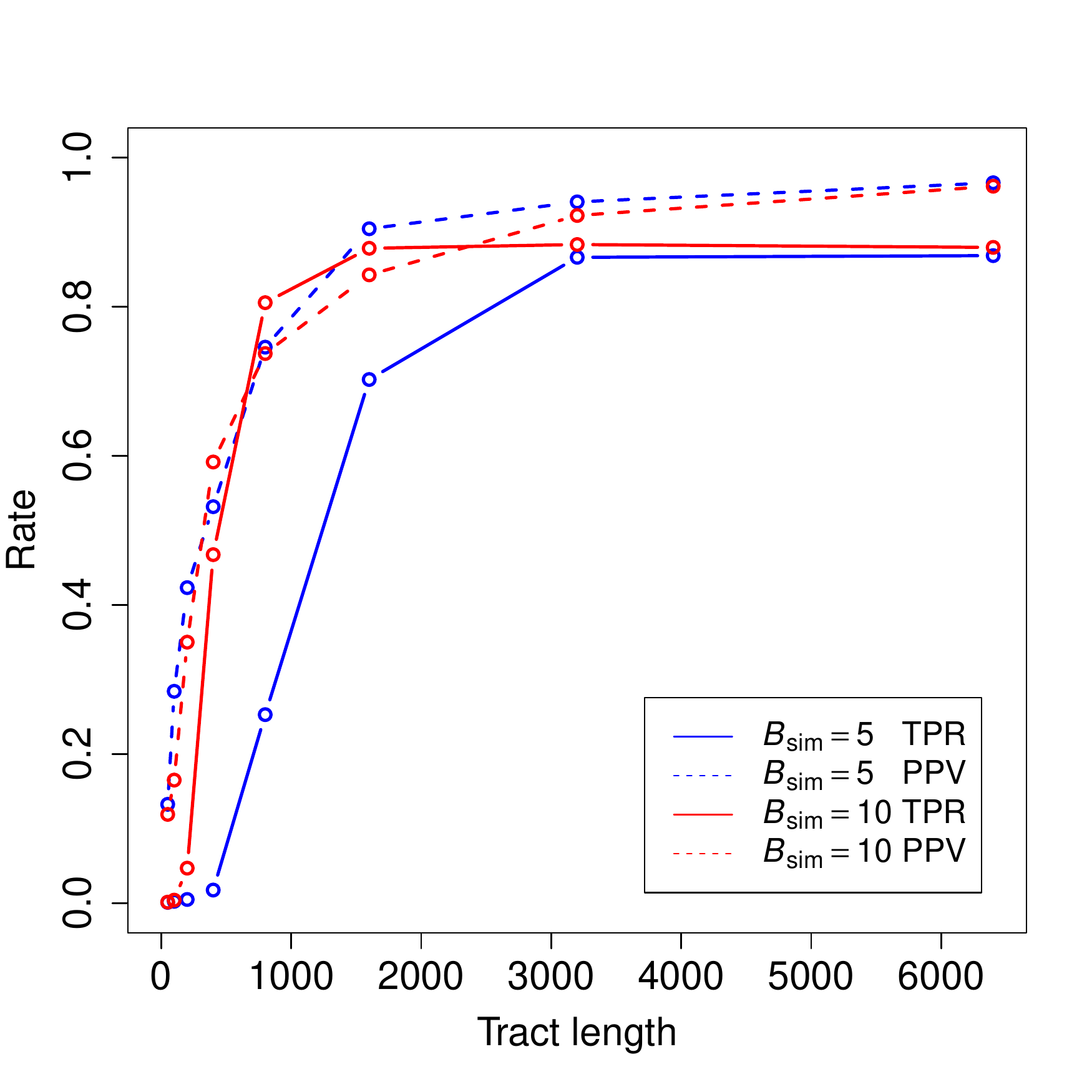}
\caption{\textbf{Power and accuracy for simulated data}.  The plot shows
  true positive rates (TPR; fraction of true gBGC bases correctly
  predicted) and positive predictive values (PPV; fraction of predicted
  bases in true gBGC tracts) as a function of tract length.  Results are
  shown for two sets of simulations, one assuming strong BGC
  ($B_{\text{sim}}=10$), and the other assuming weaker BGC
  ($B_{\text{sim}}=5$) (see Methods).  Both the power (as measured by the
  TPR) and the accuracy (as measured by PPV) of gBGC detection depend
  strongly on tract length.  At shorter lengths ($<$3000 bp) power also
  depends strongly on the strength of gBGC, while accuracy does not.  Both
  TPR and PPV are fairly high ($\sim$80\% or more) for tracts of $\geq$1 kb
  that have experienced strong gBGC, and for tracts of $\geq$1.6 kb that
  have experienced weaker gBGC.} 
\label{fig:possim}
\end{center}
\end{figure}

Given what is currently known about gBGC, it is impossible to obtain direct
measurements of the completeness and accuracy of our predicted tracts. Our simulation experiments suggest that both sensitivity and
specificity are reasonably good for tracts at least 1--2 kb in length
with $B \geq 5$, but we often miss shorter or less biased gBGC tracts
(Figure~\ref{fig:possim}), and the true distributions of tract lengths and
$B$ values are unknown (although estimates of $B=1.3$
\cite{SPENETAL06} and $B=8.7$ \cite{GALTETAL09} have been reported for
highly recombining regions).  It is important to bear in mind that $B$
represents an average along an entire branch of the phylogeny.  Many
regions may have experienced quite strong gBGC but for short evolutionary
intervals, resulting in small average values of $B$ and poor detection
power.  Thus, while our genome-wide predictions improve on what is
currently available, it seems plausible that they still represent the ``tip
of the iceberg''---a relatively small subset of all genomic regions
significantly influenced by gBGC, perhaps unusual for their length or GC-disparity.  

It is worthwhile to consider two other indirect sources of information
about our power for gBGC tract prediction.  First, Katzman et al.\
\cite{KATZETAL11} found that about 20\% of the 40 kb genomic intervals they
examined show significant \ws{} DAF skew.  If we conservatively assume one
1--2 kb tract per gBGC-influenced window, this observation would imply that
at least 0.5--1.0\% of the human genome has been influenced by gBGC on
population genomic time scales, compared with the phastBias estimate (for
$B=3$) of 0.3\%.  Second, using a method optimized for the analysis of
individual HARs, Kostka et al.\ \cite{KOSTETAL12} estimated that 24\% of
HARs experienced significant gBGC (19\% exclusively and 5\% in combination
with positive selection), or 3.7 times as many as overlap our phastBias
predictions (6.4\%).  Thus, these two imperfect indicators of power suggest
that, with $B=3$, phastBias underpredicts gBGC tracts by a factor
of roughly 2--4 or more.  The genomic coverage of our $B=2$ predictions may be
closer to the truth (1.1\%; Table \ref{tab:hmmTractStats}), but these predictions
appeared to be of poorer quality on inspection, apparently because
the phylo-HMM states with and without gBGC were insufficiently distinct to
control false positive rates.

It is also worth noting that, while the likelihood ratio tests of Kostka et al.\
\cite{KOSTETAL12} appeared to have greater power for gBGC in HARs overall,
phastBias sometimes achieves improved sensitivity by considering the entire
genome (including flanking sequences) rather than just a designated
collection of elements.  Indeed, of the thirteen HARs that overlap one of
our gBGC tracts, three were not identified by Kostka et al., apparently
for this reason.  These instances of improved sensitivity are especially
noteworthy given that phastBias must address the more difficult problem of
unconstrained genome-wide prediction, with the attendant potential for
large numbers of false positives predictions.

In principle, gBGC can overcome purifying selection and help to drive
deleterious alleles to high frequencies
\cite{GALTETAL09,DUREGALT09,GLEM10}, but it has been difficult to find
direct empirical evidence for a reduction in fitness (genetic load) caused
by gBGC.  Our predicted gBGC tracts provide two hints of an empirical link
between gBGC and fitness reductions.  First, the predicted tracts are
significantly enriched for disease-associated polymorphisms in current
human populations, suggesting that gBGC has helped to drive at least some
of these alleles to appreciable frequencies, and, indeed, may still be
active in maintaining these deleterious alleles.  Second, the inferred \ws{} substitutions in
the tracts are enriched for overlap with evolutionarily conserved sequences
in eutherian mammals, suggesting long-term purifying selection against the
(presumably) gBGC-driven G/C allele at some of these positions.  However,
other lines of evidence were inconsistent with these findings, and despite
several attempts (Supplementary Material),
we were unable to obtain reliable estimates of the overall contribution of
gBGC to deleterious substitutions in the human genome.
This estimation problem is difficult because many correlates of gBGC are
also correlates of nucleotide substitution rates, and because most
gBGC-induced substitutions are probably neutral, with only a small fraction
being deleterious.  This will be an interesting and
important topic for future work.

\clearpage
\section*{Methods}

\subsection*{Probabilistic Model}
\begin{figure}[h!]
\begin{center}
\includegraphics[width=3.25in]{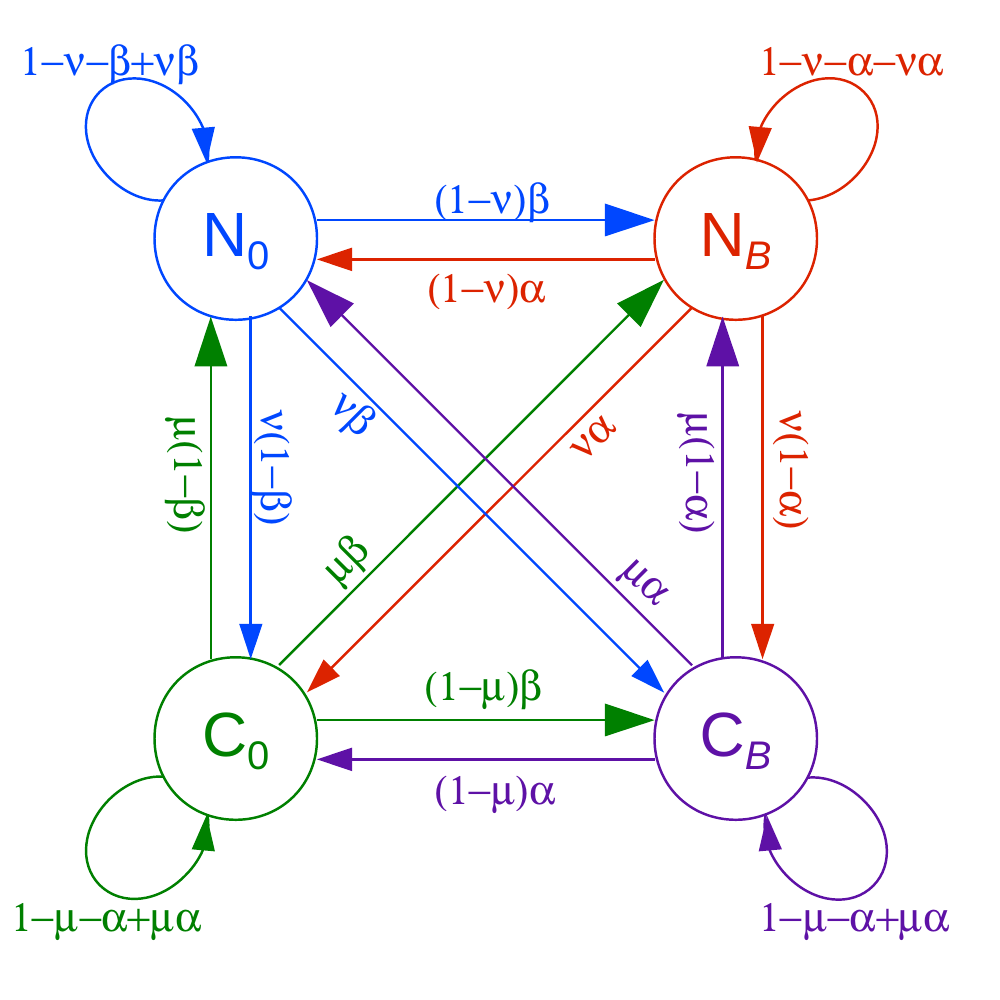}
\caption{\textbf{Phylogenetic hidden Markov model used by phastBias.}  The
  model consists of four states: neutral evolution with no 
  gBGC (N$_0$), neutral evolution with gBGC (N$_B$), evolutionary
  conservation with no gBGC (C$_0$), and evolutionary conservation with
  gBGC (C$_B$).  gBGC is assumed to influence nucleotide substitution rates
  and patterns only on the lineage leading to a designated target genome
  (human or chimpanzee in this study).  The model generalizes the phylo-HMM
  used by phastCons for prediction of evolutionarily conserved elements
  \cite{SIEPETAL05}.  The state transition probabilities 
  are defined by four parameters, denoted $\mu$, $\nu$, $\alpha$, and
  $\beta$.  See Methods and Table~\ref{tab:hmmParameters} for
  details.}
\label{fig:hmm}
\end{center}
\end{figure}

Our phylogenetic hidden Markov model has four states: one that assumes both
evolutionary conservation and gBGC (C$_B$), a second with gBGC but no
conservation (N$_B$), a third with conservation but no gBGC (C$_0$), and a
fourth with neither conservation nor gBGC (N$_0$) (Figure
\ref{fig:hmm}).  To avoid over-parameterization, we make the following
simplifying assumptions.  First, we model gBGC only on the lineage leading
to a pre-defined ``target'' genome (human or chimpanzee), because gBGC is
expected to be a transient phenomenon, typically affecting a single lineage
in any genomic position of interest.  gBGC tracts are allowed to occur on
other lineages, but these tracts are expected to have a negligible
influence on inferences in the target genome and are not directly
modeled. Second, negative selection, in contrast to gBGC, is assumed to
apply uniformly across all branches of the phylogeny.  Third, positive
selection is ignored.  We omit positive selection and lineage-specific
negative selection from the model because they are expected to be fairly
rare, to leave a relatively weak signal in the data at human-chimpanzee
evolutionary distances \cite{SIEPETAL06}, and to primarily operate at a somewhat
different genomic scale from gBGC (e.g., at the level of individual binding
sites or clusters of amino acids, rather than genomic tracts of hundreds or
thousands of bases).  We expect our modeling framework to be robust to
occasional sequences under positive or lineage-specific selection, because
the primary signal for tract prediction is a \ws{} substitution bias, and
selection generally will not produce such a bias consistently across many
bases.  Finally, we assume that the strength of gBGC and the strength of
negative selection in the target genome are constant across the genome.  A
similar homogeneity assumption is employed in phastCons and appears to have
a minimal impact on power and accuracy for element identification
\cite{SIEPETAL05}.

With these assumptions, the phylogenetic models for the four states are
defined as follows.  
\begin{enumerate}
\item{Neutral/No gBGC (N$_0$)}: Neutral evolution is described by an HKY
  substitution model \cite{HASEETAL85}, with free parameters for the
  transition/transversion ratio ($\kappa$) and stationary nucleotide
  frequencies ($\vect \pi$).  We assume the accepted tree topology for the
  species under consideration: (((human, chimpanzee), orangutan), rhesus
  macaque).  The branch length proportions were obtained from the
  Conservation tracks in the UCSC Genome Browser (assembly hg18)
  \cite{MEYEETAL13}.  (They were originally estimated from fourfold
  degenerate sites in protein coding genes under a strand-symmetric general
  reversible model.)  These branches were scaled locally to accommodate
  regional variation in mutation rate (see below).
\item{Neutral/gBGC (N$_B$)}: This model is identical to the neutral model
  except that it assumes gBGC influences substitution rates and patterns on
  the lineage leading to the target species (human or chimpanzee) according
  to the model of Kostka et al.\ \cite{KOSTETAL12}.  The strength of gBGC
  is described by the GC-disparity parameter $B>0$, which increases the
  rate of \ws{} substitutions and decreases the rate of \sw{}
  substitutions.
\item{Conserved/No gBGC (C$_0$)}: Evolutionary conservation is modeled, as
  in phastCons, by multiplying the branch lengths of the neutral model by a
  factor $\rho$ ($ 0 \le \rho \le 1$).  In all other respects, this model
  is identical to the neutral model.  Note that the application of a
  multiplicative scale factor to the neutral branch lengths is
  mathematically equivalent to selecting a constant value for the
  population-scaled selection coefficient $S$ in the joint BGC/selection
  model of Kostka et al.\ \cite{KOSTETAL12}.
\item{Conserved/gBGC (C$_B$)}: This model is identical to model N$_B$
  except that it assumes gBGC acts with strength $B$ on the lineage leading
  to the target species.
\end{enumerate} 

The state-transition probabilities are defined by four parameters, denoted
$\mu$, $\nu$, $\alpha$, and $\beta$ (Figure~\ref{fig:hmm},
Table~\ref{tab:hmmParameters}).  The parameters $\mu$ and $\nu$ are
inherited from phastCons \cite{SIEPETAL05} and describe the conditional
probabilities of transitioning from a conserved state to a neutral state,
and from a neutral state to a conserved state, respectively.  The
parameters $\alpha$ and $\beta$ are analogous, defining the conditional
probabilities of transitioning out of, and into, a gBGC tract, respectively.  The sixteen possible state
transition probabilities are obtained by multiplying the appropriate pairs
of conditional 
probabilities and enforcing the standard normalization constraints
(Figure~\ref{fig:hmm}).  This ``cross-product'' construction corresponds 
to a prior assumption of independence for the two types of transitions
(conservation $\leftrightarrow$ no conservation and gBGC $\leftrightarrow$
no gBGC). 

Given a multiple sequence alignment, standard algorithms for statistical
phylogenetics and hidden Markov models can be used to calculate the
likelihood of the data under this model, to predict the most likely state
path (Viterbi), or to calculate the marginal posterior probability of each
state at each alignment column (reviewed in \cite{SIEPHAUS05}).

\subsection*{Parameter Estimation}

In principle, the nine free parameters in our model (Table
\ref{tab:hmmParameters}) could all be estimated directly from the data by
maximum likelihood, using an expectation maximization or numerical
optimization algorithm.  In practice, however, parameter estimation is
difficult because there are no validated gBGC tracts to use for supervised
training of the model, and the signal in the data is not sufficiently
strong to support a fully unsupervised estimation procedure.  Instead, we
partition the parameters into three groups: those for the neutral
substitution process, those for the model of conserved elements, and those
specific to the gBGC tracts.  The first two groups of parameters are
pre-estimated from the data without consideration of gBGC, by what can be
considered an empirical Bayes approach.  The parameters in the third group
are then estimated by a combination of methods.

Specifically, the free parameters for the neutral substitution process
($\lambda$, $\vect \pi$, and $\kappa$) are estimated per alignment block
(see below) using phyloFit \cite{HUBIETAL11}, after conditioning on the
tree topology and branch-length proportions (as described above).  This
strategy assumes that conserved elements and gBGC tracts are sparse and
have at most a minor effect on average substitution rates for large genomic
blocks.  The three additional parameters that describe conserved elements ($\rho$,
$\mu$, and $\nu$) are inherited directly from phastCons and therefore were
simply set to the values used for the Conservation tracks in the UCSC
Genome Browser.  The remaining parameters include the GC-disparity $B$ and
the gBGC transition probabilities $\alpha$ and $\beta$.  As discussed in
the Results section, we found that $\alpha$---which can be interpreted as
an inverse prior expected length for gBGC tracts---has only a weak
influence on our predictions (within a reasonable range) and decided to
simply fix it at 1/1000, corresponding to a prior expectation of 1 kb
tracts.  We treated $B$ as a ``tuning'' parameter and considered various
possible values in a plausible range.  The final parameter,
$\beta$, was estimated from the data (separately for each alignment block)
by expectation maximization, conditional on fixed values of all other
parameters.

\subsection*{Tract Prediction}

To predict gBGC tracts based on our model, we computed marginal posterior
probabilities for the four model states at each genomic position using the
forward/backward algorithm.  We then computed the marginal posterior
probability of gBGC by summing the probabilities for states N$_B$ and
C$_B$, and we predicted tracts by applying a threshold of 0.5 to this
probability (i.e., the predicted tracts are maximal segments in which every
position has a posterior probability of at least 50\% of gBGC).  We settled on this
strategy after discovering that the more conventional Viterbi algorithm
performed poorly in this setting, evidently due to uncertainty about the
endpoints of tracts.  This uncertainty causes the probability mass for a
putative gBGC tract to be distributed across many possible HMM state paths,
and as a result, the Viterbi algorithm often fails to predict a tract even
when the posterior probability of gBGC is close to one.  A potential
drawback of our thresholding strategy is that fluctuating posterior
probabilities could lead to highly fragmented tract predictions.  However,
we found that the posterior probability function was quite
smooth in practice (probably owing to small values of the state transition
probabilities) and fragmentation was not a problem.  For example, at $B=3$,
only about 2\% of the predicted human tracts fall within 50 base pairs of
another tract. Nonetheless, when analyzing the genomic distribution of gBGC tracts relative to one another and to telomeres, we merged adjacent tracts (within 1 kb) in order to reduce any bias introduced by over fragmentation (Supplementary Material).

\subsection*{Genome-wide Alignments and Preprocessing}

Our analyses of both simulated and real data were based on genome-wide
alignments obtained from the UCSC Genome Browser
(\url{http://genome.ucsc.edu}) \cite{MEYEETAL13}.  We began with the 44-way
vertebrate alignments produced with multiz \cite{BLANETAL04} (hg18
assembly) and extracted the rows corresponding to the human, chimpanzee,
orangutan, and rhesus macaque genomes, discarding alignment columns
containing only gaps in these sequences.  We also discarded columns in
which the human genome contained a gap.
Human-referenced alignments were used for both the human and chimpanzee
gBGC tract predictions, as chimpanzee-based multiple alignments are not
available.  For convenience in processing, the resulting four-way
alignments were partitioned into blocks of approximately 10 megabases (Mb)
in length.  The boundaries between blocks were required to occur in regions
uninformative about gBGC (due to $>$1~kb with lack of alignment with the
other species).  

\subsection*{Simulation Study}
We simulated human sequences with gBGC tracts for each 10 Mb block in the
real genome-wide alignments as follows.  First, we identified positions at
which any sequence contained a CpG dinucleotide, because substitution rates
are likely to be substantially elevated at such sites.  Next, we used
phastCons to identify conserved elements in the four species.  We then
fitted a phylogenetic model to the alignment columns in each of four
categories (neutral/non-CpG, conserved/non-CpG, neutral/CpG, conserved/CpG)
by estimating $\kappa$, $\vect \pi$, and $\lambda$ for the most data-rich
category (neutral/non-CpG), then estimating a separate $\lambda$ for
the CpG category (using phyloFit) and applying a branch-length scale-factor
of 0.3 to the 
conserved categories.  Next, we defined an alternative ``gBGC'' instance of
each of the four estimated models by modifying the substitution rate matrix
for the human branch according to our model of gBGC~\cite{KOSTETAL12} and a
given choice of $B$ (here denoted $B_{\text{sim}}$).  In this way, we
obtained eight phylogenetic models, representing all combinations of
conservation / no conservation, CpG / no CpG, and gBGC / no gBGC.

We generated synthetic human sequences by assigning one of these eight
models to each alignment column, as follows.  The conservation and CpG
status of each column was maintained as originally annotated, so that the
synthetic alignments would resemble the original ones as much as possible.
The gBGC status was set to ``no gBGC'' for most columns, but set to
``gBGC'' for tracts of fixed size at randomly selected locations, at an
average gBGC coverage of 0.1\%.  We then simulated a new human base for
each alignment column conditional on the assigned phylogenetic model and the
observed chimpanzee, orangutan, and rhesus macaque bases.  This was
accomplished using the `postprob.msa' function in RPHAST, which computes the
marginal distribution over bases at any node in the phylogeny conditional
on a given phylogenetic model and collection of observed bases, using the
sum-product algorithm.  This
function computes the desired distribution for the human base if the human
sequence is masked and treated as missing data in the input.  A particular
base was selected by sampling from this marginal distribution.

We performed this simulation procedure for combinations of $B_{\text{sim}}
\in \{3,5,10\}$ and fixed tract lengths of 200, 400, 800, 1600, 3200, and
6400.  For each set of simulated alignments, we predicted gBGC tracts as
described in the previous section, assuming several different values for
the tuning parameter $B$.  For each data set and value of $B$, we
calculated the true positive rate (number of correctly predicted gBGC bases
/ total number of gBGC bases), false positive rate (number of incorrectly
predicted gBGC bases / total number of non-gBGC bases), and positive
predictive value (number of correctly predicted gBGC bases / number of
predicted gBGC bases).

\subsection*{Genomic Annotations}

We compared the predicted gBGC tracts with exon and intron definitions from
Gencode version 3c and Ensembl genes~\cite{FLICETAL11}, and with
annotations of lincRNAs, miRNAs, miscRNAs, small non-coding RNAs, NMD
transcripts, and pseudogenes from Gencode version 14~\cite{HARRETAL12}. We
also compared them with LINE and SINE elements from the rmskRM327 table in
the UCSC Table Browser~\cite{KAROETAL04}, and with a set of high-confidence
predictions of transcription factor binding sites based on ChIP-seq data
from ENCODE~\cite{ENCOCONS12} (L.\ Arbiza, I.\ Gronau, B.A.\ Aksoy, M.J.\
Hubisz, B.\ Gulko, A.\ Keinan, and A.\ Siepel, in revision).  In addition,
we compared the tracts with genome-wide recombination rate estimates from
the 1000 Genomes Project \cite{1KGCONS10}, recombination hotspots from the
October 2006 release of HapMap~\cite{HAPMCONS07}, and chimpanzee
recombination rate estimates from the PanMap project~\cite{AUTOETAL12}.

Disease-associated SNPs were obtained from several sources. SNPs annotated
with ``pathogenic'' or ``probable pathogenic'' clinical significance were
downloaded on October, 25, 2011 from dbSNP~\cite{SHERETAL01}. The HGMD
dSNPs were obtained from the Supplementary Material of
reference~\cite{NECSETAL11}.  Regions of the human genome with positive 
genetic associations with disease were taken from the Genetic Association
Database~\cite{ZHANETAL10} on February 2, 2012. The level of evidence for
the function of non-coding SNPs was downloaded from the
RegulomeDB~\cite{BOYLETAL12} web site on December 12, 2012.  All data not
in reference to the GRCh36/hg18 assembly were mapped to hg18 using the
`liftOver' tool from the UCSC Genome Browser.

\subsection*{Control Regions}

To evaluate the statistical significance of various properties of interest,
we compared the predicted gBGC tracts with sets of control regions matched
to them in number, length distribution, and chromosome assignment.  We also
ensured that the control regions were matched to the gBGC tracts by GC
content (by stratifying predictions and controls into 100 bins), which
is known to correlate strongly with several relevant genomic 
features.  We obtained a null distribution for each statistic of interest
(such as the number of tracts overlapping exons, or the number human tracts
overlapping orthologous chimpanzee tracts), by computing a value of the
statistic for each of 1000 randomly sampled replicates of the control
regions.  One-sided empirical $p$-values were computed as the fraction of
sampled control sets for which the statistic was at least as extreme as
observed in the predicted tracts.  As noted in the text, we occasionally
considered alternative 
sets of control regions designed to accommodate known biases in genomic
regions of interest. For example, when evaluating the significance of
overlap with disease-associated SNPs from HGMD and dbSNP, we used control
regions matched to the predicted tracts in terms of their degree of exon
overlap, since these sets consist mostly of coding SNPs. Similarly, for
RegulomeDB, which is focused on non-coding SNPs, we used control regions
that matched the overlap of the gBGC tracts with the set of SNPs considered
by RegulomeDB.

\subsection*{Analysis of Derived Allele Frequencies}

Our analysis of derived allele frequences was based on genotype data from
the low-coverage pilot data set from the 1000 Genomes Project released in
July 2010~\cite{1KGCONS10}. These comprise SNP calls for the 22 autosomes
in three HapMap population panels: YRI (59 individuals), CEU (60
individuals), and CHB-JPT (60 individuals). We computed the \ws{} DAF skew
of all gBGC tract SNPs as normalized $U$ values from a Mann-Whitney $U$ test
on the derived allele frequencies of \ws{} and \sw{} SNPs, as previously
described \cite{KATZETAL11}. A \ws{} DAF skew of 0.5 indicates no bias, and
values greater than 0.5 indicate that \ws{} mutations are favored.

\subsection*{Purifying Selection at Orthologous Positions}
We evaluated evidence of purifying selection at regions of other mammalian
genomes by considering (1) evolutionary conservation scores
for mammals and (2) patterns of polymorphism in chimpanzee.  In both cases, we
compared regions orthologous to the predicted human tracts and regions
orthologous to control regions.  The first comparison used phyloP
CONACC conservation scores \cite{POLLETAL09} at positions of human-specific
\ws{} substitutions within the predicted tracts.  To avoid possible biases
from human-specific substitutions or from the chimpanzee sequence used to
assign these substitutions to the human branch, we re-computed the phyloP
scores based on alignments of eutherian mammals from which the human and
chimpanzee sequences had been removed (leaving 30 mammalian species).  We
compared the tracts with both exon- and GC-matched control regions
(Supplementary Figure~\ref{fig:phyloP}).  For the polymorphism
analysis, we used data from the PanMap Project \cite{AUTOETAL12} and
considered all nucleotides within the predicted tracts, because the subset
of sites with both human-specific \ws{} substitutions and chimpanzee
polymorphisms was very small (42 sites).

\section*{Acknowledgments}

We thank Laurent Duret, Graham Coop, and members of the Pollard and Siepel
research groups for helpful discussions.

\subsection*{Funding}

Funding was provided by National Institutes of Health (NIGMS) grant GM82901
(to K.S.P.\ and A.S.), a David and Lucile Packard Fellowship for Science
and Engineering (to A.S.), and a PhRMA Foundation Informatics Fellowship (to J.A.C.).

\enlargethispage{\baselineskip}

\clearpage
\bibliography{gbgc_tract_arXiv}

\clearpage
\section*{Tables}

\begin{table}[ht]
\caption{Summary of HMM parameters}
\label{tab:hmmParameters}
\begin{minipage}{6in}
\vspace{1ex}
\begin{small}
\begin{tabular}{cclc}
\hline
\bf{Parameter} & \bf{Group}\footnote{neut = parameters for neutral phylogenetic
  model,  cons = parameters for conserved elements (inherited from
  phastCons), gBGC = parameters for gBGC tracts.} & \bf{Description}  & \bf{Value}\\
\hline
$\lambda$ & neut &
Scale factor for neutral branch lengths & estimated per
10 Mb block\\ 
$\vect \pi$\footnote{Multivariate parameter (three degrees of freedom).} & neut & Equilibrium nucleotide frequencies & estimated per
10 Mb block\\ 
$\kappa$ & neut & Transition/transversion ratio & estimated per
10 Mb block\\ 
\hline
$\rho$ & cons & Branch length scale factor in conserved state &
0.3\footnote{Values used for the Conservation tracks
  in the UCSC Genome Browser (see \cite{SIEPETAL05}).}\\
$\mu$ & cons& Transition prob.\ conserved$\rightarrow$neutral & 0.022$^\textit{c}$ \\
$\nu$ & cons & Transition prob.\ neutral$\rightarrow$conserved & 0.0095$^\textit{c}$\\
\hline
$B$ & gBGC & GC-disparity (strength of gBGC) & 2, 3\footnote{Value used for primary analyses.}, 4, 5, 10 \\
$\alpha$ & gBGC & Transition prob.\ gBGC$\rightarrow$non-gBGC &
0.001\footnote{Corresponds to prior expected length of 1 kb.}\\ 
$\beta$ & gBGC & Transition prob.\ non-gBGC$\rightarrow$gBGC & optimized by EM\\
\hline
\end{tabular}
\end{small}
\end{minipage}
\end{table}

\vspace{2in}

\begin{table}[ht]
\begin{center}
\begin{minipage}{6in}
\begin{center}
\caption{Summary of predicted gBGC tracts}
\label{tab:hmmTractStats}
\vspace{1ex}
\begin{tabular}{lrrrrr}
\hline
\bf{Species} & \bf{{\em B}} & \bf{Number} & \bf{Coverage} & \bf{Mean length} & \bf{Median length} \\
\hline
Human & 2 & 12362 & 1.103\% & 2567 & 1008	\\
Human & 3 & 9439 & 0.334\% & 1018 & 788	\\
Human & 4 & 7712 & 0.217\% & 810 & 628 \\
Human & 5 & 6750 & 0.157\% & 670 & 514 \\
Human & 10 & 5210 & 0.073\% & 400 & 276	\\
\hline
Chimpanzee & 3 & 8677 & 0.252\% & 841 & 663 \\
Chimpanzee & 10 & 7062 & 0.068\% & 278 & 198 \\
\hline
\end{tabular}
\end{center}
\end{minipage}
\end{center}
\end{table}

\begin{table}[ht]
\begin{center}
\begin{minipage}{6in}
\begin{center}
\caption{Recombination rates in gBGC tracts}
\label{tab:human_chimp_recomb}
\vspace{1ex}
\begin{tabular}{llll}
\hline
\bf{Recombination} & \bf{Human gBGC tract} & \bf{Chimpanzee gBGC tract} &  \bf{GC-matched Control} \\
\bf{Map} & \bf{Rate (cM /Mb)} & \bf{Rate (cM /Mb)} & \bf{Rate (cM /Mb)} \\
\hline
Human & 3.85 & 1.81\footnote{Obtained by mapping chimpanzee tracts to
  orthologous positions in the human genome.} &  1.61 \\
Chimpanzee  & 1.33\footnote{Obtained by mapping human tracts to orthologous
  positions in the chimpanzee genome.} & 1.71 &  0.78 \\
\hline
\end{tabular}
\end{center}
\end{minipage}
\end{center}
\end{table}

\begin{table}[ht]
\begin{center}
\begin{minipage}{6in}
\begin{center}
\caption{Enrichment for disease-associated regions}
\label{tab:disease_enrich}
\vspace{1ex}
\begin{tabular}{lrrr}
\hline
\bf{Disease-associated} & \bf{gBGC Tract} & \bf{Avg. Control} &   \\
\bf{Region Set} & \bf{Overlap} & \bf{Overlap} & \bf{{\em p}-value} \\
\hline
dbSNP Pathogenic &113 & 46.3 & 0.005 \\
HGMD & 346 & 178.2 & 0.031 \\
RegulomeDB (classes 1--5) & 26474 & 20768.4 & $<$0.001 \\
GAD & 485 & 419.7 & $<$0.001 \\
\hline
\end{tabular}
\end{center}
\end{minipage}
\end{center}
\end{table}

\clearpage
\section*{Figures}

\clearpage

\setcounter{figure}{0}
\setcounter{table}{0}

\makeatletter 
\renewcommand{\thefigure}{S\@arabic\c@figure} 
\renewcommand{\thetable}{S\@arabic\c@table} 
\renewcommand{\thesection}{S\@arabic\c@section} 
\makeatother

\section*{Supplementary Methods}
\subsection*{Clustering, recombination rate, and distance to telomeres}

To investigate the degree to which the predicted gBGC tracts are
clustered, we calculated for each 
gBGC tract the distance to the nearest other gBGC tract
(distance-to-nearest). One potential caveat with this type of analysis is
that gBGC tracts were annotated by thresholding posterior probabilities in
our HMM
(see Methods), which might cause predictions to occur very close together 
simply because of fluctuations in posterior probability along the
genome. Therefore, for these analyses, we merged gBGC tracts with 
their neighbors if the distance between them was less than 1 kb (the
expected gBGC tract length).  

\begin{figure}

\begin{minipage}{.48\textwidth}
\textsf{A}
\vspace{-.2ex}

\includegraphics[width=.9\textwidth, trim=.1cm .5cm 1cm 1.2cm, clip=true]{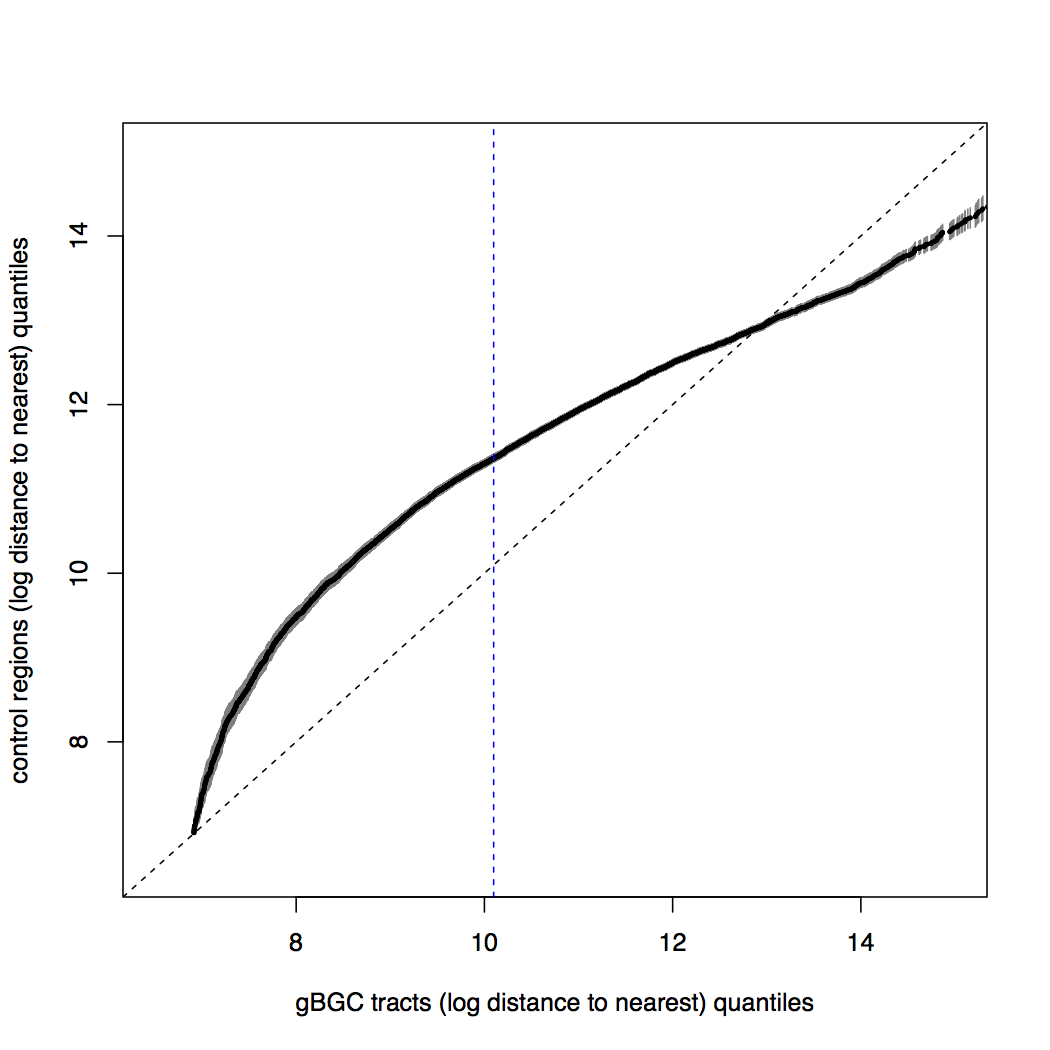}
\end{minipage}
\begin{minipage}{.48\textwidth}
\textsf{B}
\vspace{-.2ex}

\includegraphics[width=.9\textwidth, trim=.1cm .5cm 1cm 1.2cm, clip=true]{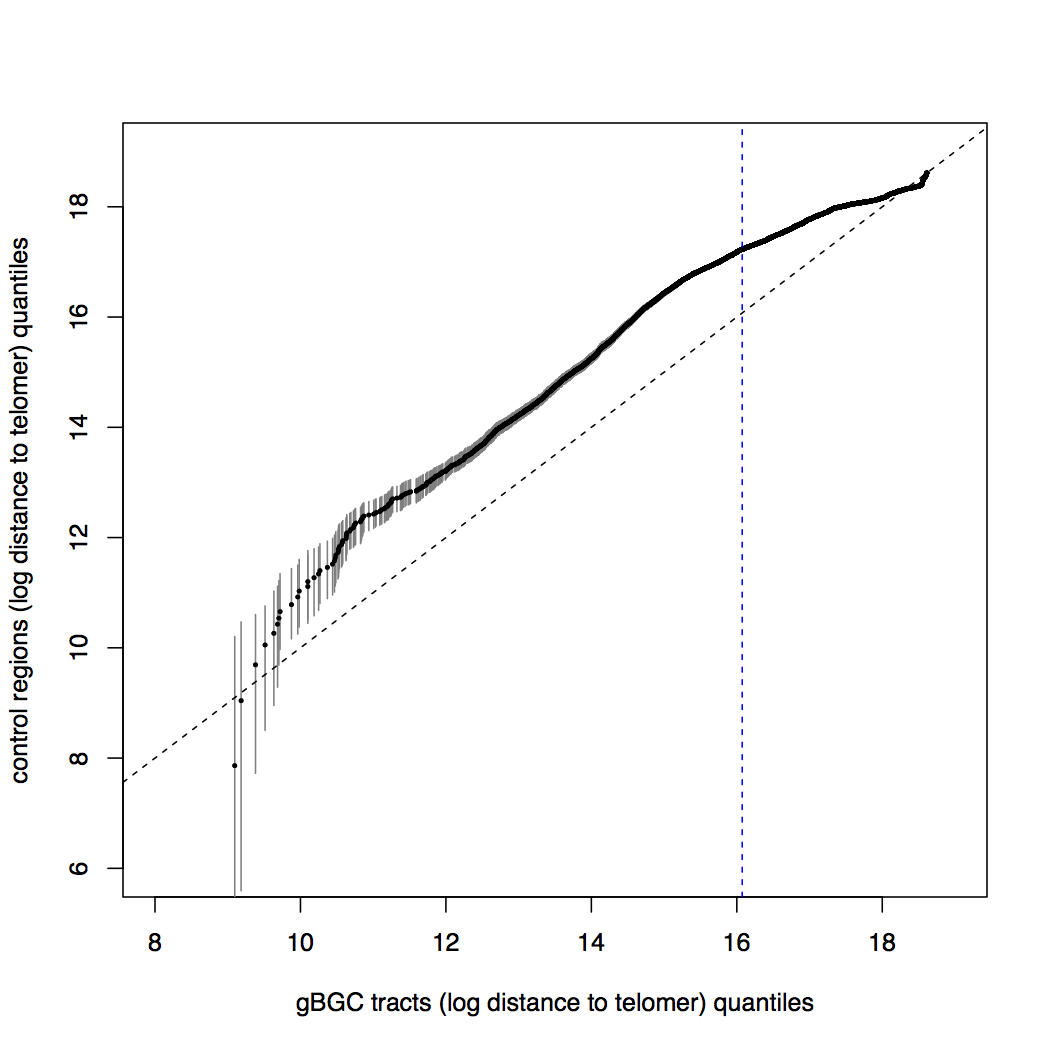}
\end{minipage}
\vspace{1ex}

\begin{minipage}{.48\textwidth}
\textsf{C}
\vspace{-.2ex}

\includegraphics[width=.9\textwidth, trim=.1cm .5cm 1cm 1.2cm, clip=true]{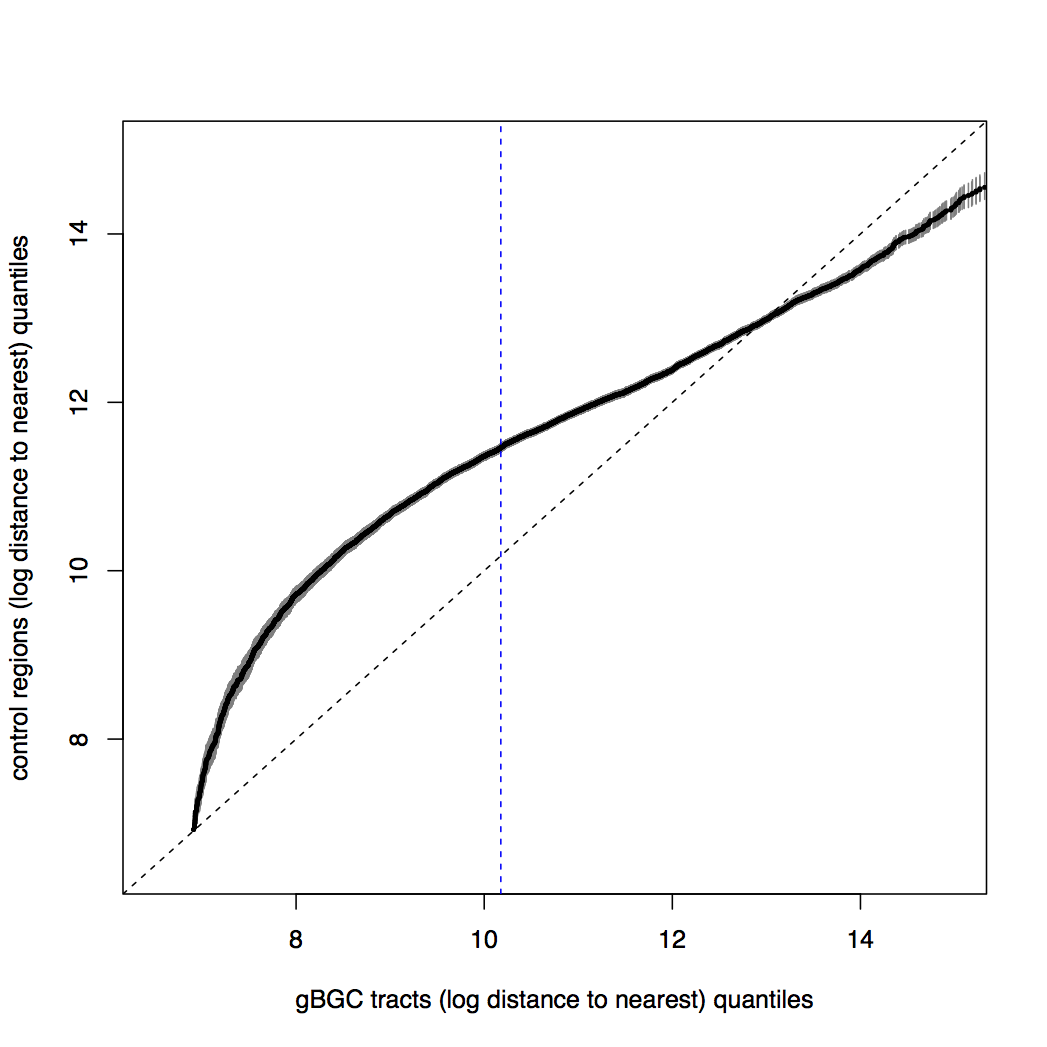}
\end{minipage}
\begin{minipage}{.48\textwidth}
\textsf{D}
\vspace{-.2ex}

\includegraphics[width=.9\textwidth, trim=.1cm .5cm 1cm 1.2cm, clip=true]{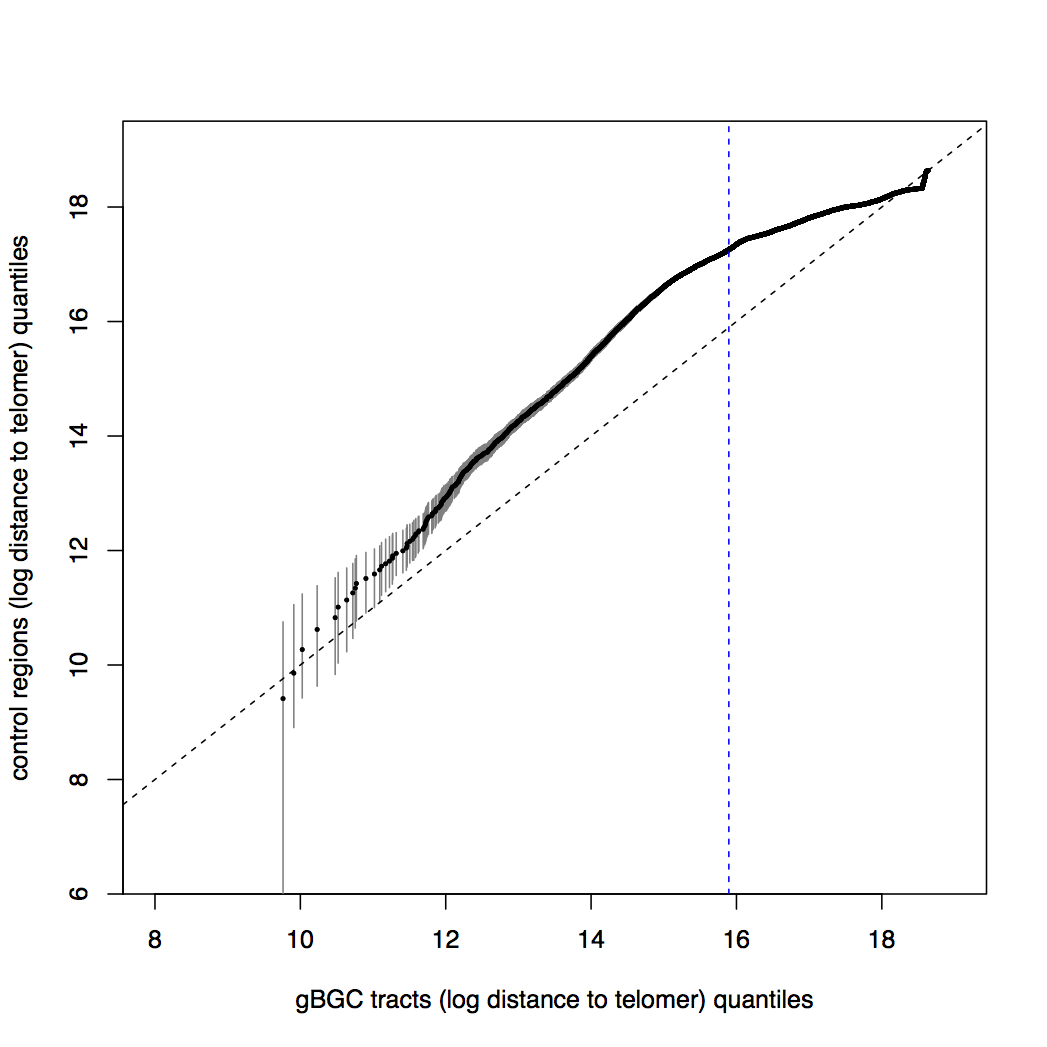}
\end{minipage}
\vspace{1ex}

\caption{
  \label{fig:qqplots} {\bf gBGC tracts are clustered and closer to
    telomeres than expected by chance.} This figure shows qq-plots
  contrasting quantiles observed in gBGC tracts ($x$-axis) with medians of
  quantiles observed in  GC-matched control sets (points,
  $y$-axis). The gray regions correspond to the data range observed across
  control regions (with the 1\% highest and 1\% lowest values removed),
  while the vertical blue dashed line denotes the median of the gBGC
  tracts. Panels A and B show these plots for distance-to-nearest and
  the distance-to-telomere in human, C and D show the corresponding
  plots for chimpanzee.  }

\end{figure}

\emph{gBGC tracts are close in human and chimp.}\hspace{.2cm} To assess the
closeness of gBGC-tracts in human and chimp we repeated the
distance-to-nearest calculation for 1,000 random GC-matched control sets
(in human and chimp separately), and then contrasted the distribution we
observed in these controls with the distribution in the gBGC-tracts. Supplementary Figure
\ref{fig:qqplots} (panels A and C) summarizes the results in terms of
quantile-quantile plots between the two distributions.  This illustrates
that gBGC-tracts are significantly closer together than regions in control
sets. For human, the median distance-to-nearest is 24,305 bp, while the
average of median nearest distances across control sets is 86,064.26 bp
(with a standard deviation of 1,571.1). For chimp, the median distance-to-nearest in the gBGC tracts is 26,286 bp, and the average median distance
across control sets is 94,882.92 bp (sd = 1,888.2). In both species, we never observe a median
distance-to-nearest in the control sets as low as in the
gBGC tracts ($z$-score based $p<\,10^{-100}$).

\emph{gBGC tracts are close to telomeres in human and chimp.}\hspace{.2cm}
To assess distance to telomeres we performed the same analysis, but used
distance to the nearest telomere in place of distance to the nearest
neighboring tract (Supplementary Figure \ref{fig:qqplots}, panels B and D). 
The median distance-to-telomere in
human and chimpanzee are 9,568,725 bp and 8,002,971 bp, while the averages
across control sets are 30,393,137 bp (sd = 329,252.1) and 31,209,994 bp
(sd = 370,981.5), respectively.  In neither
species do we observe a median distance in the control sets that is as low as
in the gBGC tracts ($p< \,10^{-100}$).

\emph{Distance-to-telomere and recombination rate partially, but not entirely,
  account for the closeness of gBGC tracts.}\hspace{.2cm} Next we asked to what extent distance-to-telomere and recombination
rate could be driving the observed proximity of gBGC tracts to one
another. To address this question we fit the following linear model to the
data for each species:
\[
E[X] = a + \beta_1 d + \beta_2 r.
\]
Here, $X$ denotes the logarithm of the 
distance to the nearest neighboring gBGC tract, $d$ is the log
distance-to-telomere, and $r$ is the mean recombination rate of a tract. We
find that, in both species, both $\beta_1$ and $\beta_2$ are significantly
different from 
zero (see Supplementary Table~\ref{tab:linmod}) and the model predicts
gBGC tracts to be close together near the telomeres and far
apart in areas 
of low recombination rate (also see Supplementary Figure~\ref{fig:boxplots}).  This is
also reflected in Spearman correlation coefficients between
distance-to-nearest and distance-to-telomere (0.36 in human, 0.49 in chimp) and recombination rate ($-$0.18 in human, $-$0.20 in chimp). Despite
these significant associations the multiple coefficient of determination
($R^2$) of the linear model is only about 13\% in human and 22\% in
chimpanzee. That is, the majority of the variance in the
distance-to-nearest observations remains unexplained when taking
distance-to-telomere and recombination rate into account.

\subsection*{Estimation of Fraction of Number of Deleterious Substitutions
  Due to gBGC}

\begin{figure}[h]
\begin{center}
\includegraphics[width=6.83in]{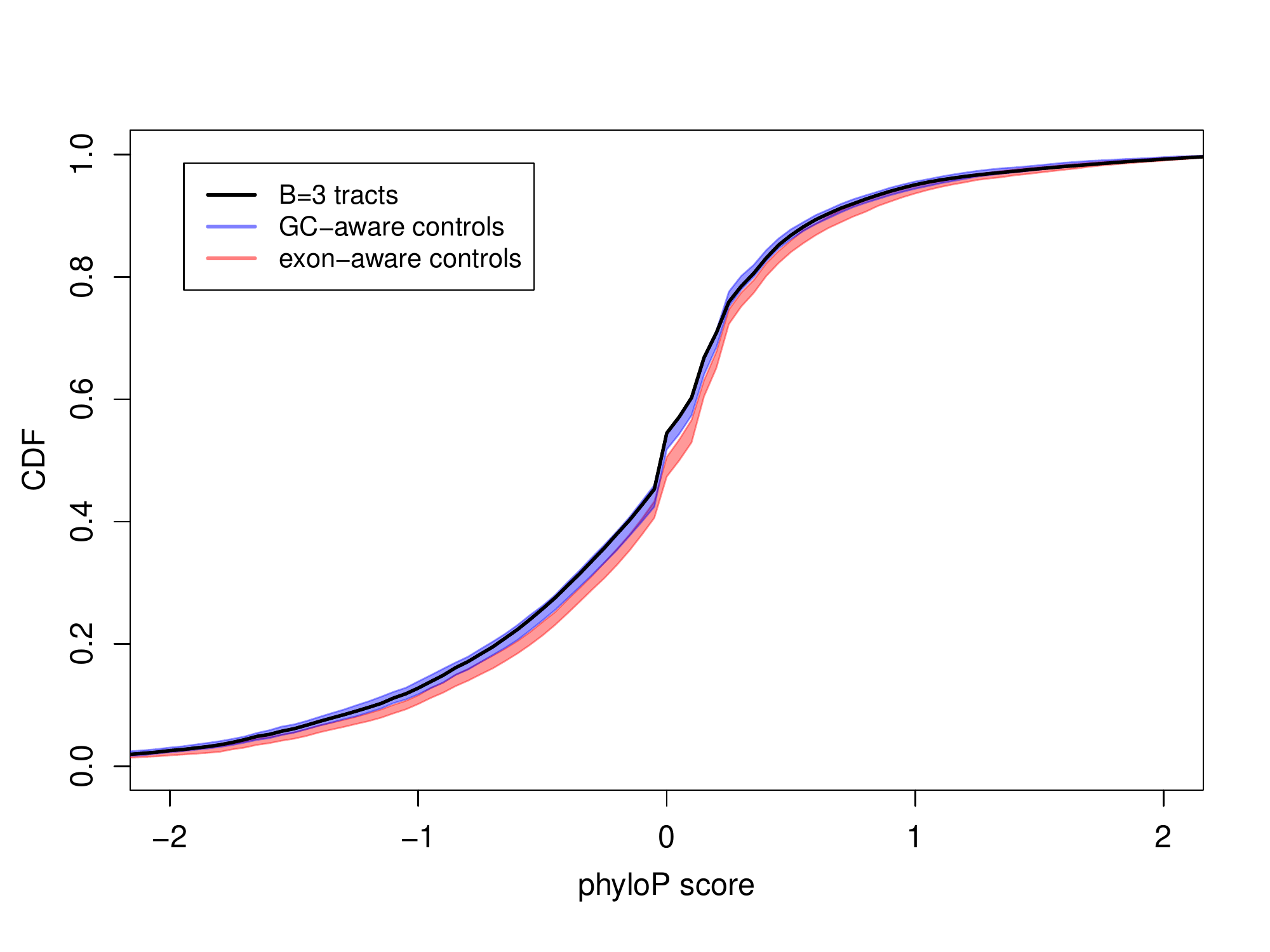}
\caption{\textbf{Conservation at sites of \ws{} substitutions within
    tracts.}  PhyloP scores were calculated at sites within the predicted
  human tracts at which \ws{} substitutions occurred on the human lineage.
  They were also calculated at sites of similar human-specific \ws{}
  substitution within the GC- and exon-matched control groups (1000
  replicates).  The scores were calculated for mammalian alignments from
  which the human and chimpanzee sequences had been removed.  A higher
  phyloP score ($x$-axis) indicates greater evolutionary
  conservation. Although there are slight differences between the
  distributions for the tracts and the control groups, there is no clear
  excess of conservation at \ws{} sites in the tracts.}
\label{fig:phyloP}
\end{center}
\end{figure}

We attempted to estimate the number of deleterious substitutions driven by
gBGC by two methods.  First, we compared the bulk distributions of
evolutionary conservation scores in regions in eutherian mammals
orthologous to the human tracts and control regions (Supplementary Figure~\
\ref{fig:phyloP}).  As described in the Methods section, we used scores
computed without the human and chimpanzee sequences to avoid biases from
human-specific substitutions.  Using various methods, we attempted to
decompose the bulk distribution of phyloP scores into a component that
could be explained by the distribution observed for control regions and an
excess of evolutionary conservation not explainable based on the control regions
(see \cite{POLLETAL09} for a description of similar methods).  Second, we
devised a method for estimating the expected excess number of human \ws{}
substitutions in the tracts, relative to the controls.  Briefly, this
method involved estimating scale factors for neutral phylogenetic models in
the non-human species at positions of human \ws{} substitutions in the
tracts and controls, and then examining the normalized difference between
these factors.  However, we found that both of these methods were highly
sensitive to the choice of control regions, as well as to technical issues
such as the choice of threshold for trimming the left-most tail of the CDF
in the mixture decomposition.  In the end, we concluded that the patterns
of evolutionary conservation were simply too similar in the tracts and
control regions, and too dependent on a complex combination of covariates
(GC-content, recombination rate, exon overlap, etc.), to lead to meaningful
estimates of the number of deleterious substitutions driven by gBGC.

%\bibliography{compbio-gbgc}

\clearpage

\section*{Supplementary Figures}
\begin{figure}[h!]
\begin{center}
\includegraphics[width=6.5in]{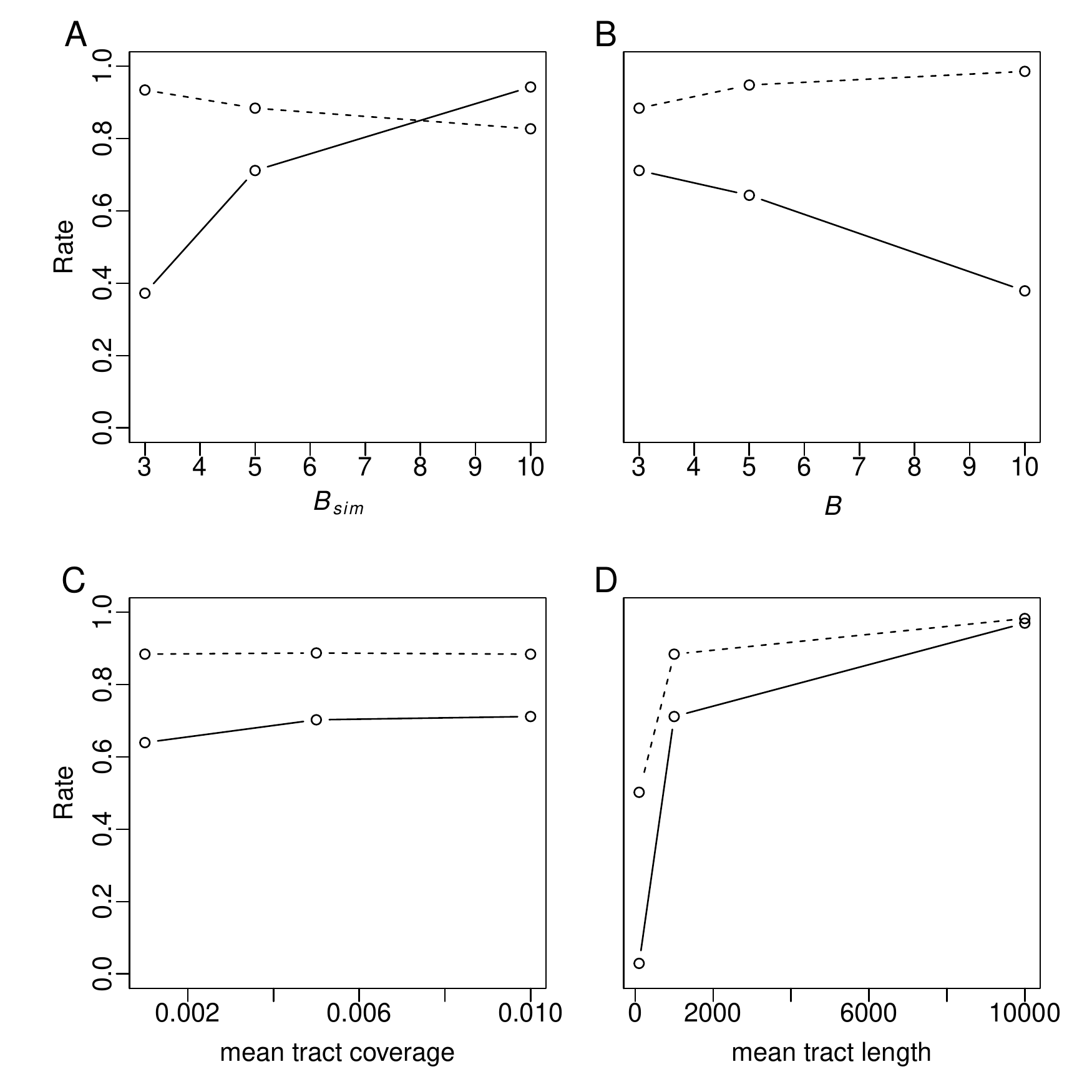}
\caption{\textbf{Additional simulation results.}  Power and accuracy for
  gBGC tract prediction as a function of (A)~gBGC strength
  ($B_{\text{sim}}$), (B)~the tuning parameter $B$, (C)~mean tract
  coverage, and (D)~mean tract length. Solid lines represent basewise true
  positive rate (TPR) and dotted lines represent positive predicted value
  (PPV).  In each plot, tracts were simulated with $B_{\text{sim}}=5$, a
  geometric length distribution with a mean of 1 kb, and mean coverage of
  1\%, unless otherwise specified by the $x$-axis.  The phylo-HMM was run
  with the same parameter settings used for the genome-wide predictions,
  including $B=3$, except in (B) (where $B$ is varied).}
\label{fig:morePosSims}
\end{center}
\end{figure}

\begin{figure}[h]
\begin{center}
\includegraphics[width=6.5in]{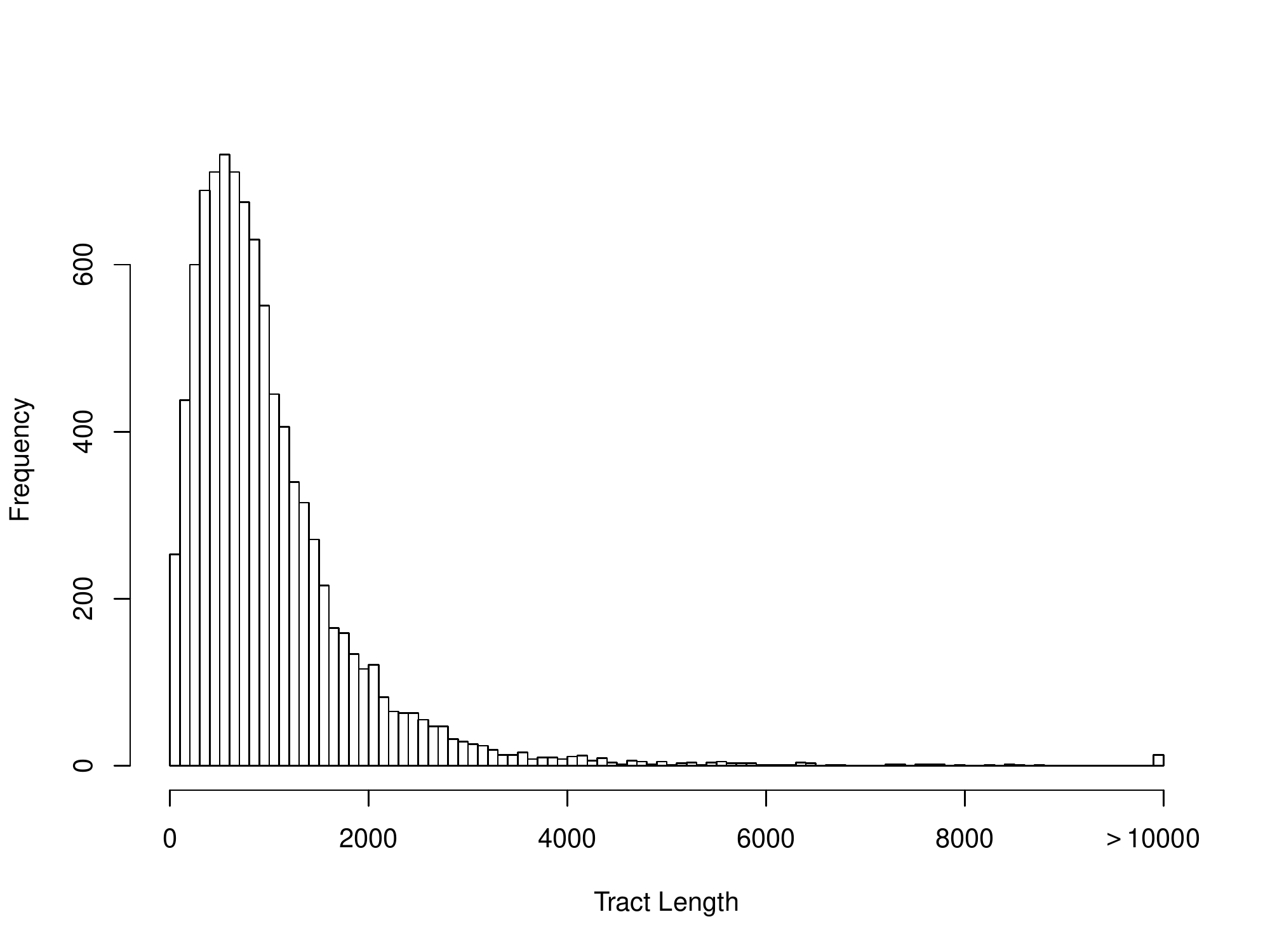}
\caption{\textbf{Length distribution of predicted human gBGC tracts for $B=3$.}}
\label{fig:length_dist}
\end{center}
\end{figure}

\begin{figure}[h]
\begin{center}
\includegraphics[width=6.5in]{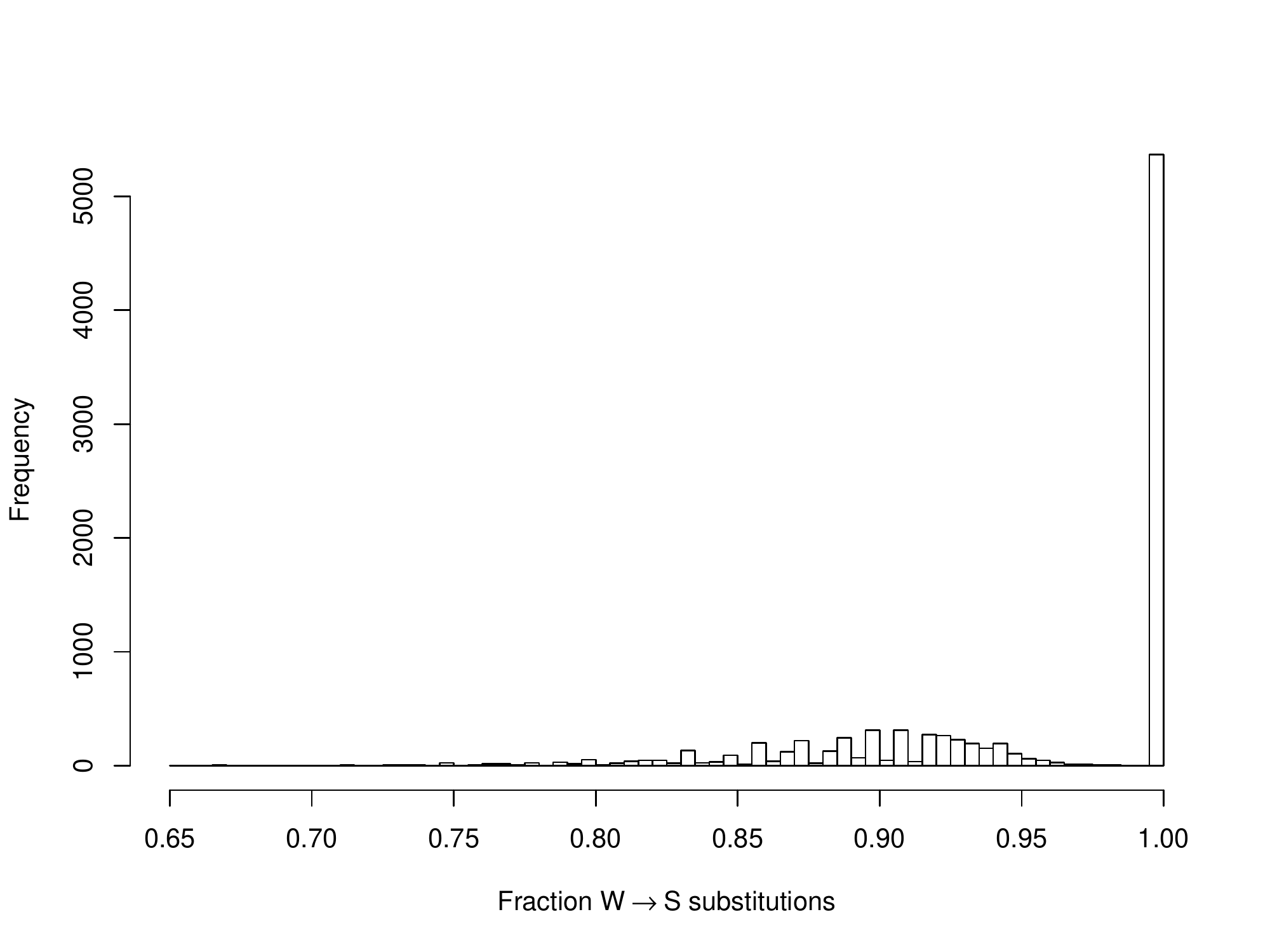}
\caption{\textbf{\ws{} bias distribution for human gBGC tracts for $B=3$.}
  Histogram of \ws{} bias, which is computed for each tract as the fraction
  of all \ws{} and \sw{} substitutions along the human lineage which are
  \ws{}.  Human-chimpanzee substitutions were polarized by assuming the
  allele observed in orangutan (ponAbe2) was ancestral.}
\label{fig:bias_dist}
\end{center}
\end{figure}

\begin{figure}[h]
\begin{center}
\includegraphics[width=6in]{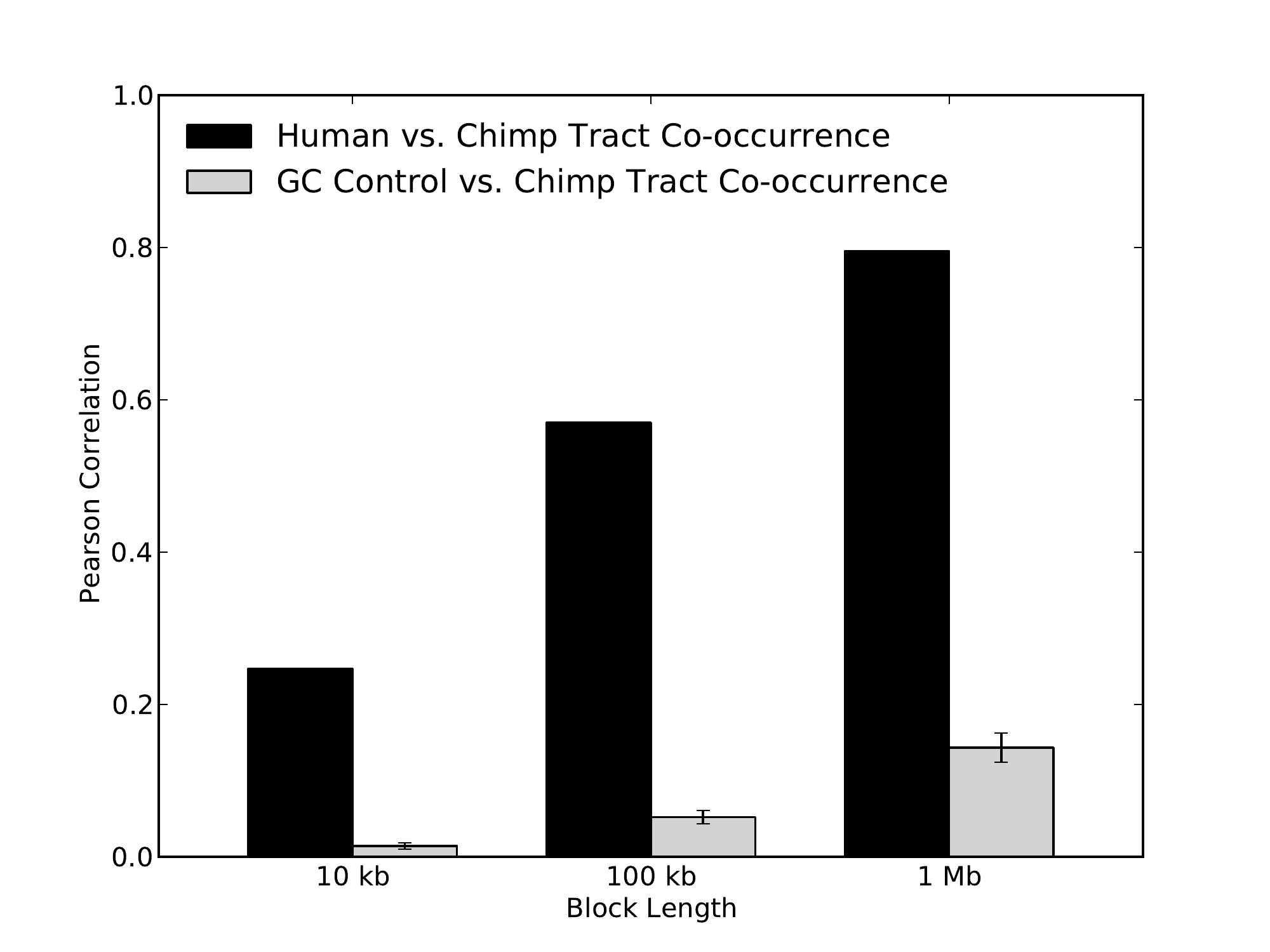}
\caption{\textbf{Human and chimpanzee gBGC tracts are found in broadly similar locations, but exhibit fine scale differences.} The fraction of bases in gBGC tracts is correlated between human and orthologous chimpanzee regions (Figure~\ref{fig:human_v_chimp}). The strength of this correlation increases as larger blocks of the genome are considered.  The gray bars give the average and standard deviation of the correlations observed between the gBGC fraction in 1000 GC-matched human control regions and the orthologous chimpanzee regions.}
\label{fig:hc_tract_correspondence}
\end{center}
\end{figure}

\begin{figure}[h]
\begin{center}
\includegraphics[width=6in]{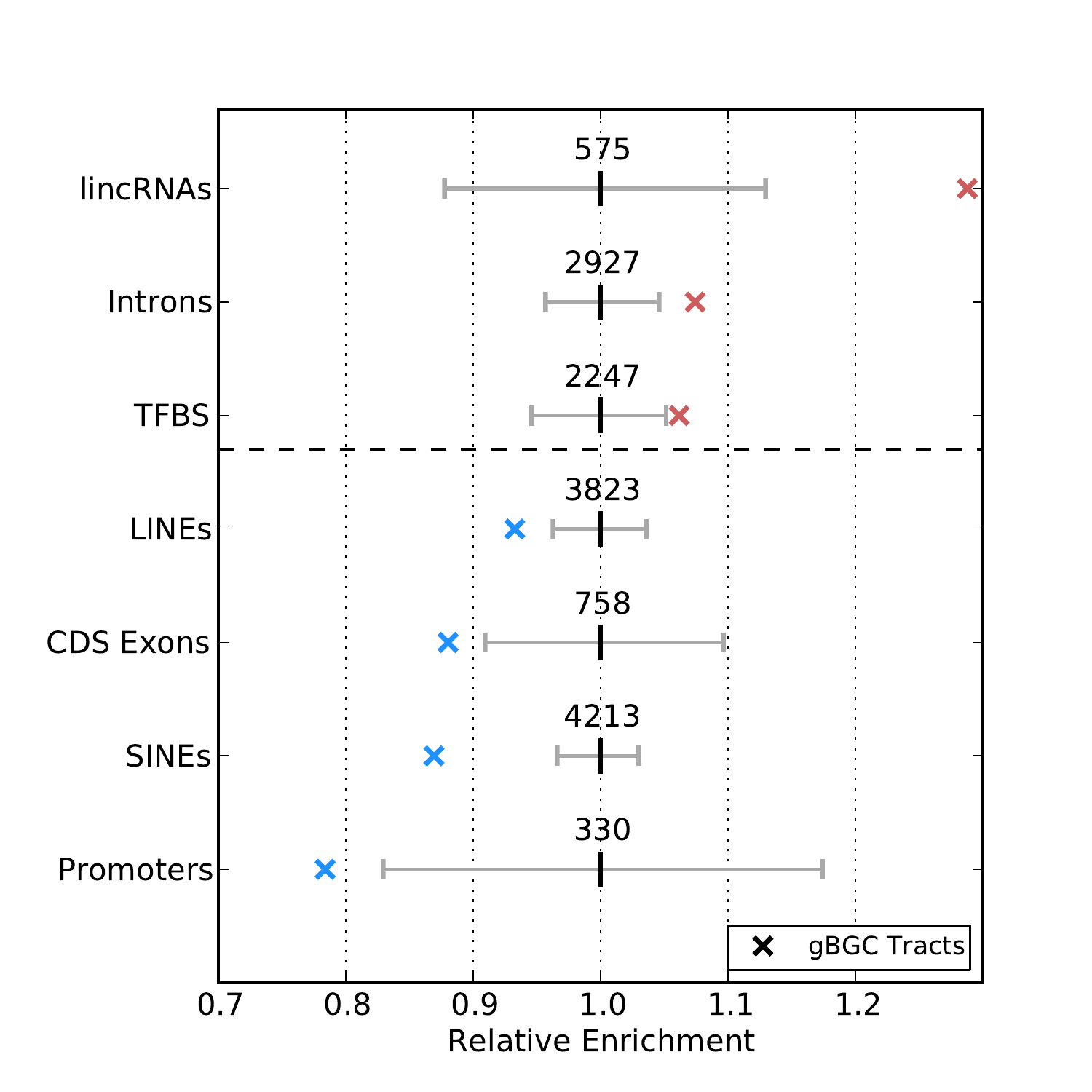}
\caption{\textbf{Genomic features significantly enriched or depeleted in
    gBGC tracts.} For each genomic feature, we compared the number of
  overlaps observed with gBGC tracts with those observed in 1000 random
  GC-matched control regions.  The gray bars give the minimum
  and maximum overlaps observed in the random sets.  Shown are all features
  that are significantly ($p < 0.05$) underrepresented (blue) or
  overrepresented (red) in the tracts. See the Methods for a full list of
  genomic features considered. Note that the tracts are more strongly
  enriched for recombination hotspots (not shown, 1.54x) and for high
  recombination rates (Table~\ref{tab:human_chimp_recomb}),
  both of which were considered separately.}
\label{fig:enriched_features}
\end{center}
\end{figure}

\begin{figure}[h]
\begin{center}
\includegraphics[width=6.83in]{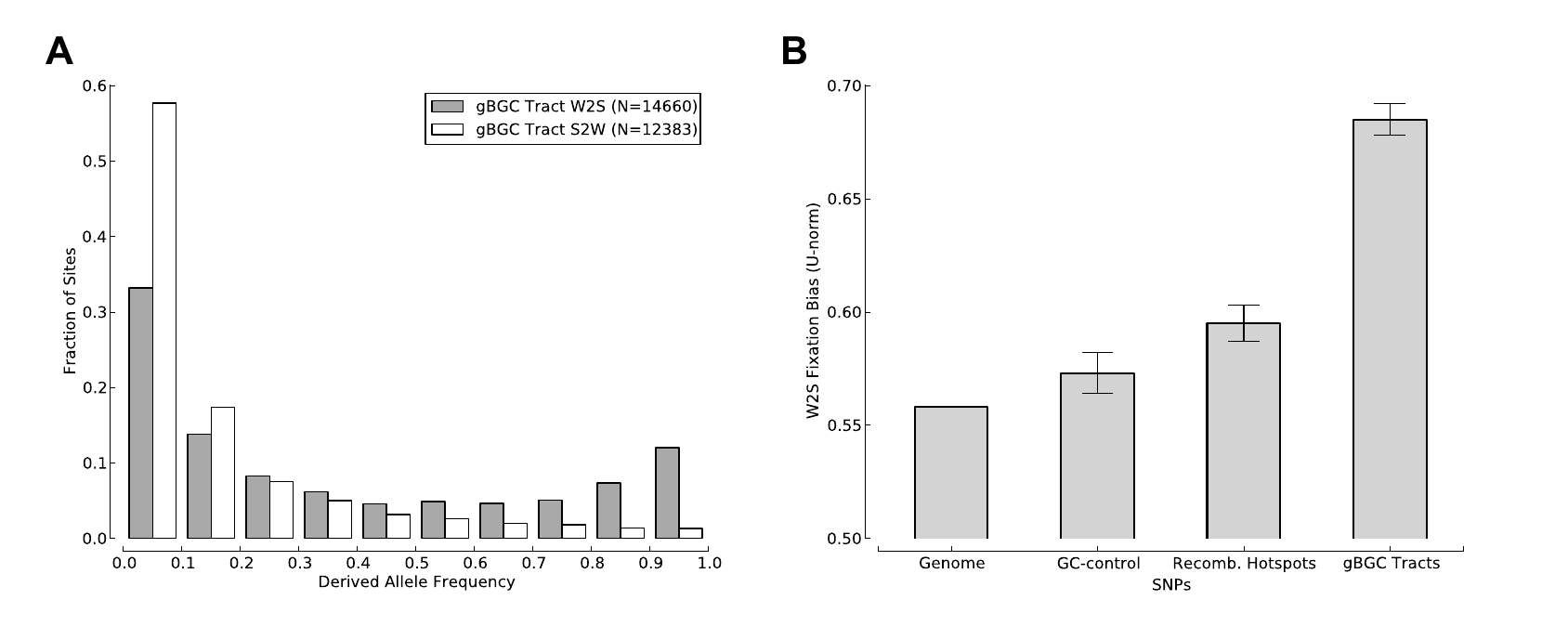}
\caption{\textbf{Human polymorphism data indicates an ongoing preference
    for the fixation of G and C alleles in the predicted gBGC tracts.}  This figure
  shows the same plots as Figure~\ref{fig:gbgc_daf_spectra}, but is based on an analysis
  in which polymorphic sites were masked from the alignments. \textbf{(A)} \ws{} changes in gBGC
  tracts have significantly higher derived allele frequencies than \sw{}
  changes.  This result was obtained on the YRI population from the 1000 Genomes Project, and patterns for other populations were similar (data not shown).  \textbf{(B)} The $U$-norm, a
  measure of the degree of \ws{} bias (see Methods), is significantly
  higher in gBGC tracts than in the entire genome or in GC-matched control
  regions.  Recombination hotspots also show somewhat
  elevated values but much less elevated than the predicted tracts.  The
  error bars indicate 95\% confidence intervals. }
\label{fig:ongoing_bias_nopoly}
\end{center}
\end{figure}

\begin{figure}[h]
\begin{center}
\includegraphics[width=6.83in]{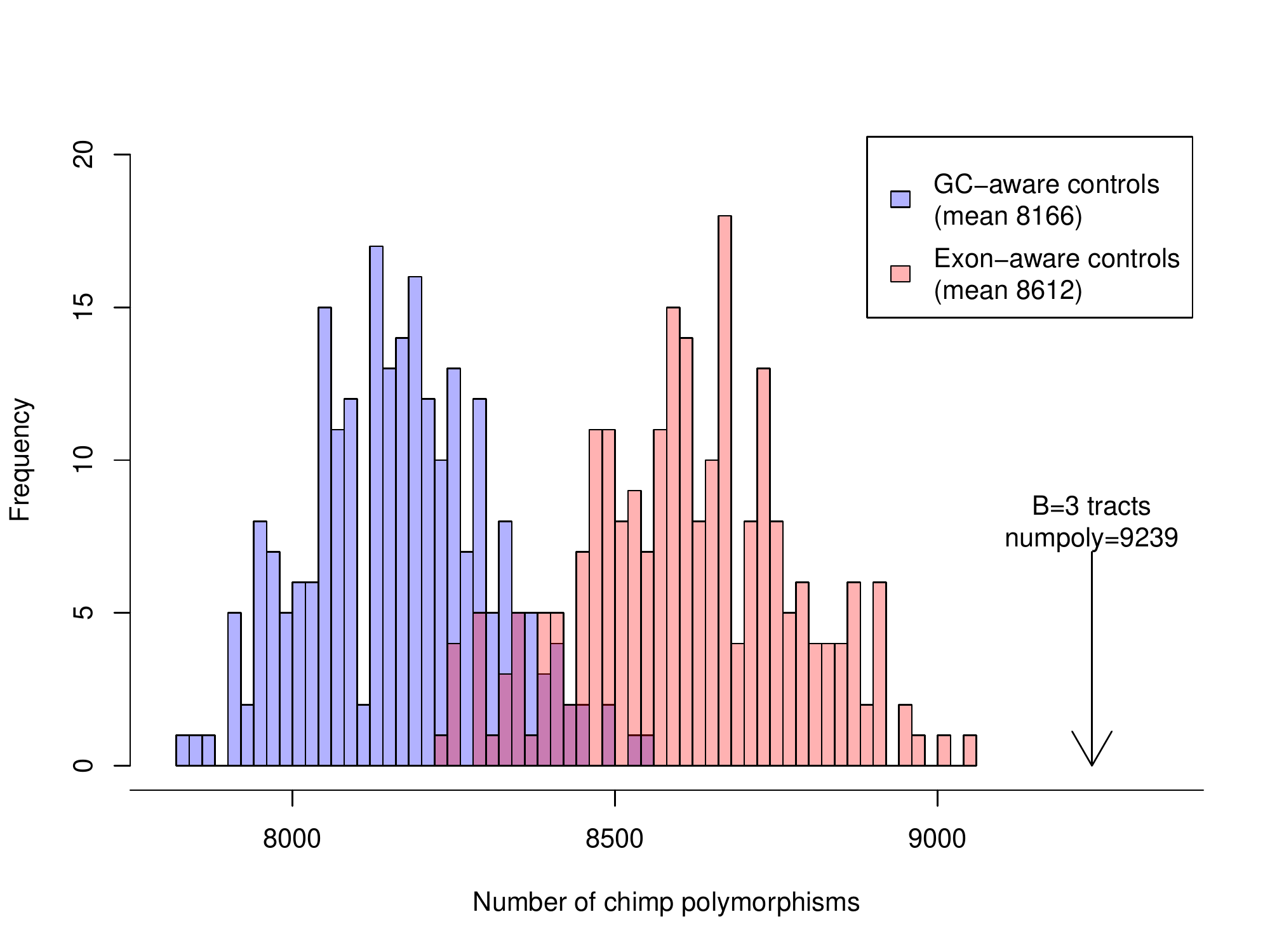}
\caption{\textbf{Number of chimpanzee polymorphisms in regions orthologous
    to the gBGC tracts, compared to controls.} We observed significantly
  more chimpanzee polymorphisms in regions orthologous to the tracts than
  those orthologous to the control groups.  This is the opposite of the
  observation that would be expected if the regions orthologous to the
  tracts were under purifying selection in the chimpanzee. }
\label{fig:chimpPolyCount}
\end{center}
\end{figure}

\begin{figure}[h]
\begin{center}
\includegraphics[width=6.83in]{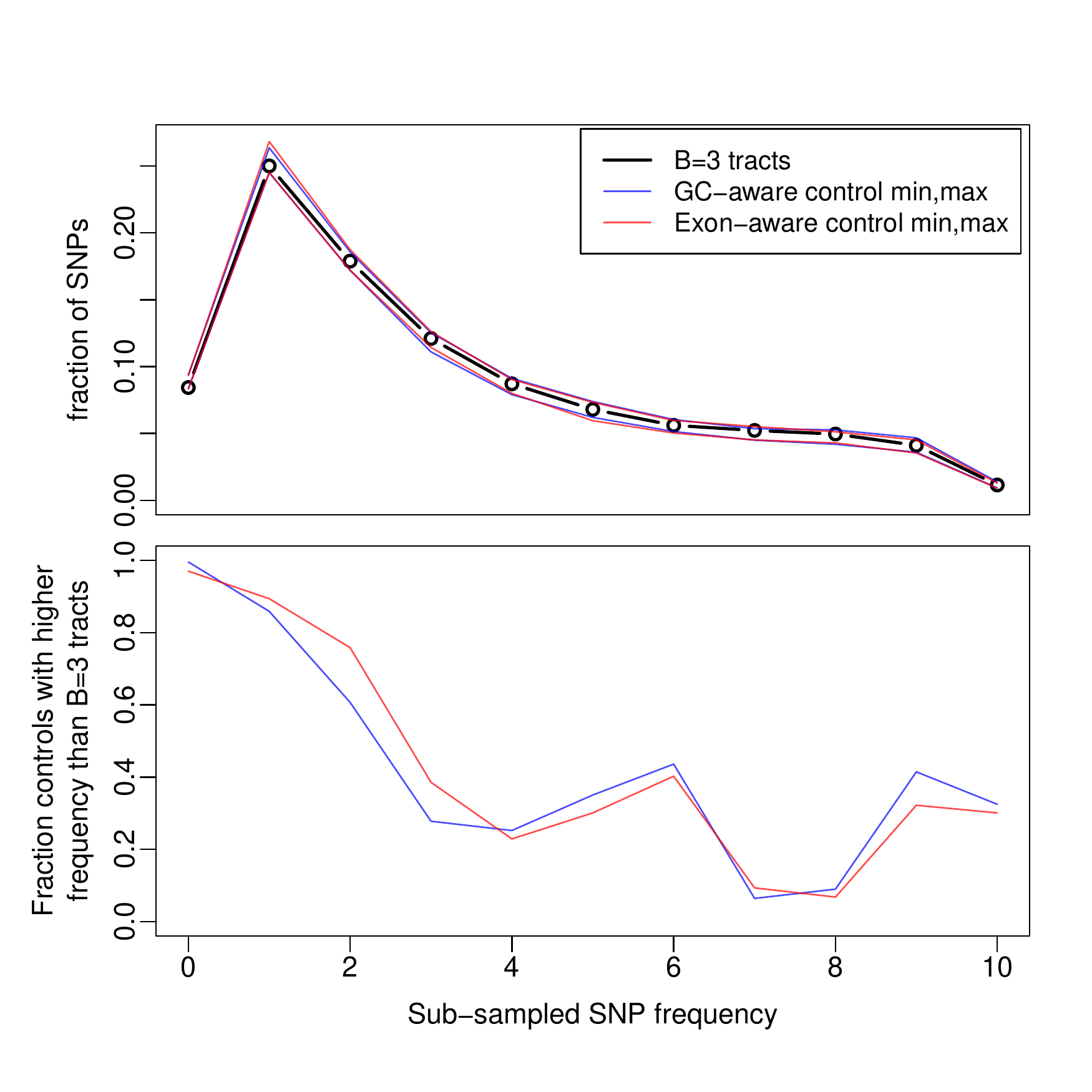}
\caption{\textbf{Derived allele frequency spectrum of chimpanzee
    polymorphisms in regions orthologous to the tracts}.  The top plot
  shows the derived allele frequency spectrum (polarized using the
  orangutan allele) for chimpanzee polymorphisms
  in regions orthologous to the $B=3$ gBGC tracts, compared with the
  minimum and maximum from 1000 control groups.  The bottom plot shows the
  fraction of samples from each control group with a higher frequency than
  observed in the real tracts.  We observe no significant excess of
  low-frequency derived alleles in regions orthologous to the tracts.}
\label{fig:chimpPolyFreq}
\end{center}
\end{figure}

\begin{figure}[h]
\begin{center}
\includegraphics[width=6.5in]{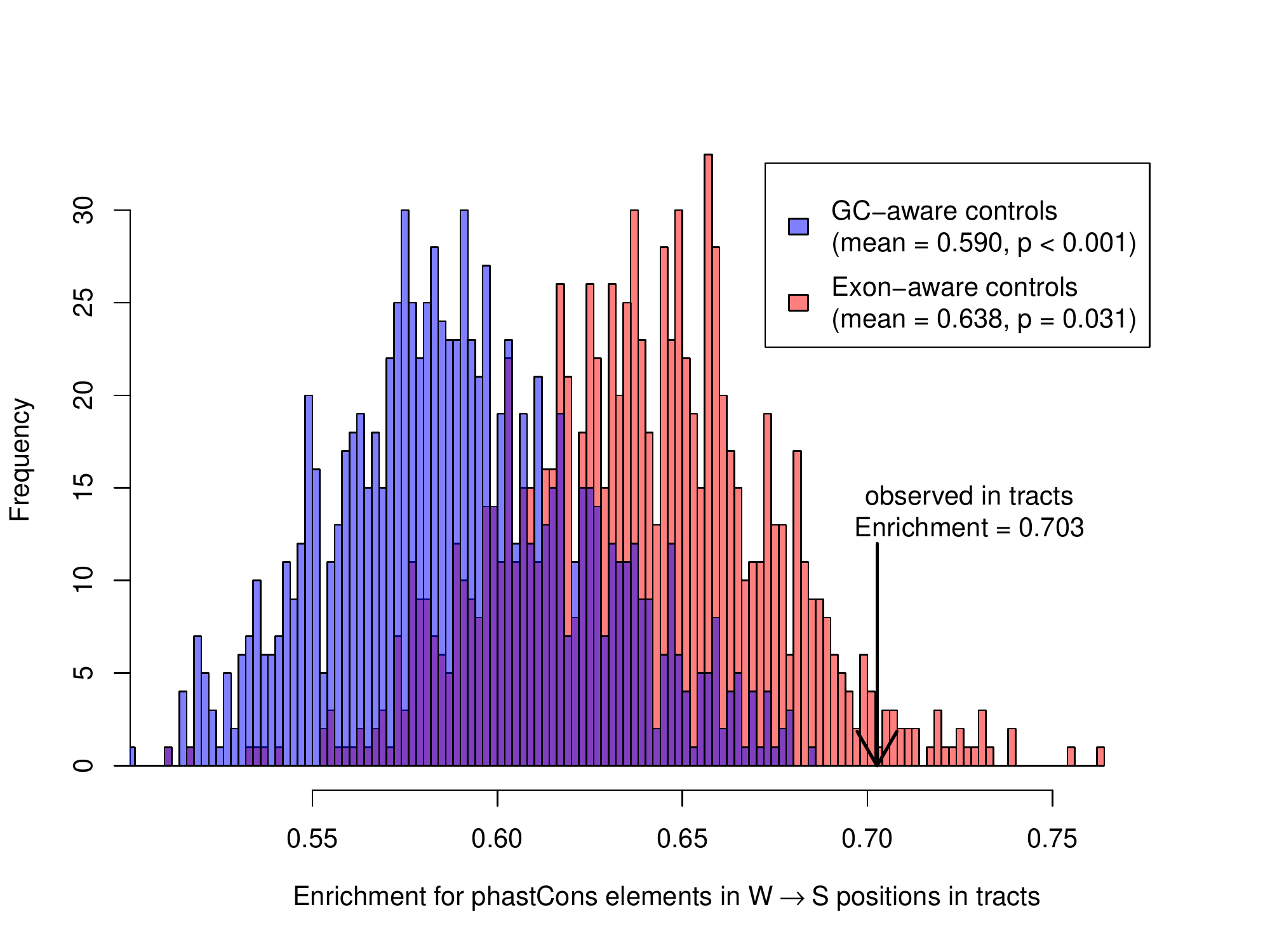}
\caption{\textbf{\ws{} sites within the predicted human tracts are enriched for phastCons elements compared to controls}.  
  Enrichments were calculated as the number of \ws{} sites within tracts falling in phastCons elements, divided by the number expected if these were distributed independently.  The histograms show enrichment in our sets of 1000 GC- and exon-aware control tracts, and the arrow shows the value observed in the $B=3$ gBGC tracts.}
\label{fig:phastConsEnrich}
\end{center}
\end{figure}

\begin{figure}

\begin{minipage}{.48\textwidth}
\textsf{A}
\vspace{-.4ex}

\includegraphics[width=.9\textwidth, trim=.1cm .5cm 1cm 1.2cm, clip=true]{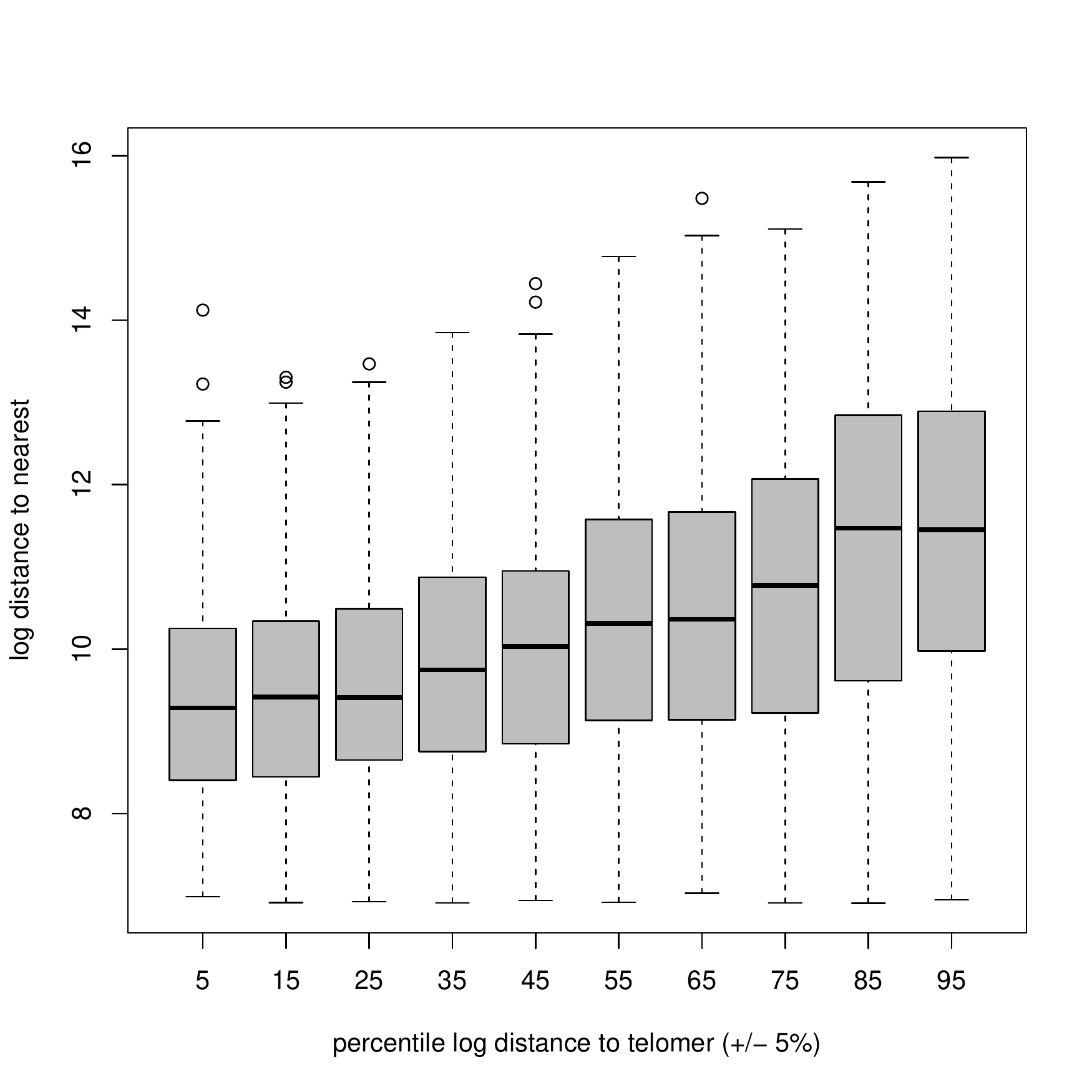}
\end{minipage}
\begin{minipage}{.48\textwidth}
\textsf{B}
\vspace{-.4ex}

\includegraphics[width=.9\textwidth, trim=.1cm .5cm 1cm 1.2cm, clip=true]{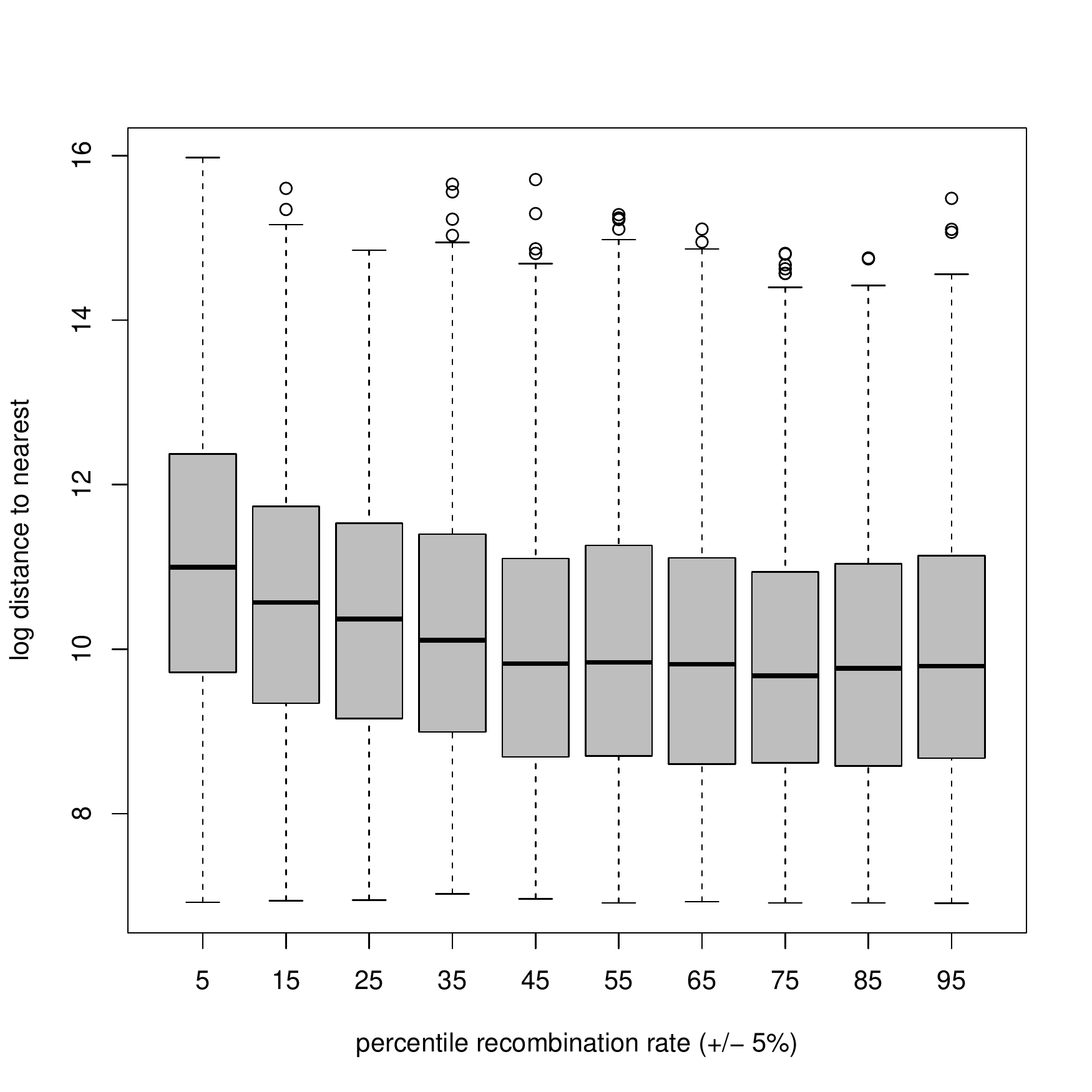}
\end{minipage}
\vspace{1ex}

\begin{minipage}{.48\textwidth}
\textsf{C}
\vspace{-.4ex}

\includegraphics[width=.9\textwidth, trim=.1cm .5cm 1cm 1.2cm, clip=true]{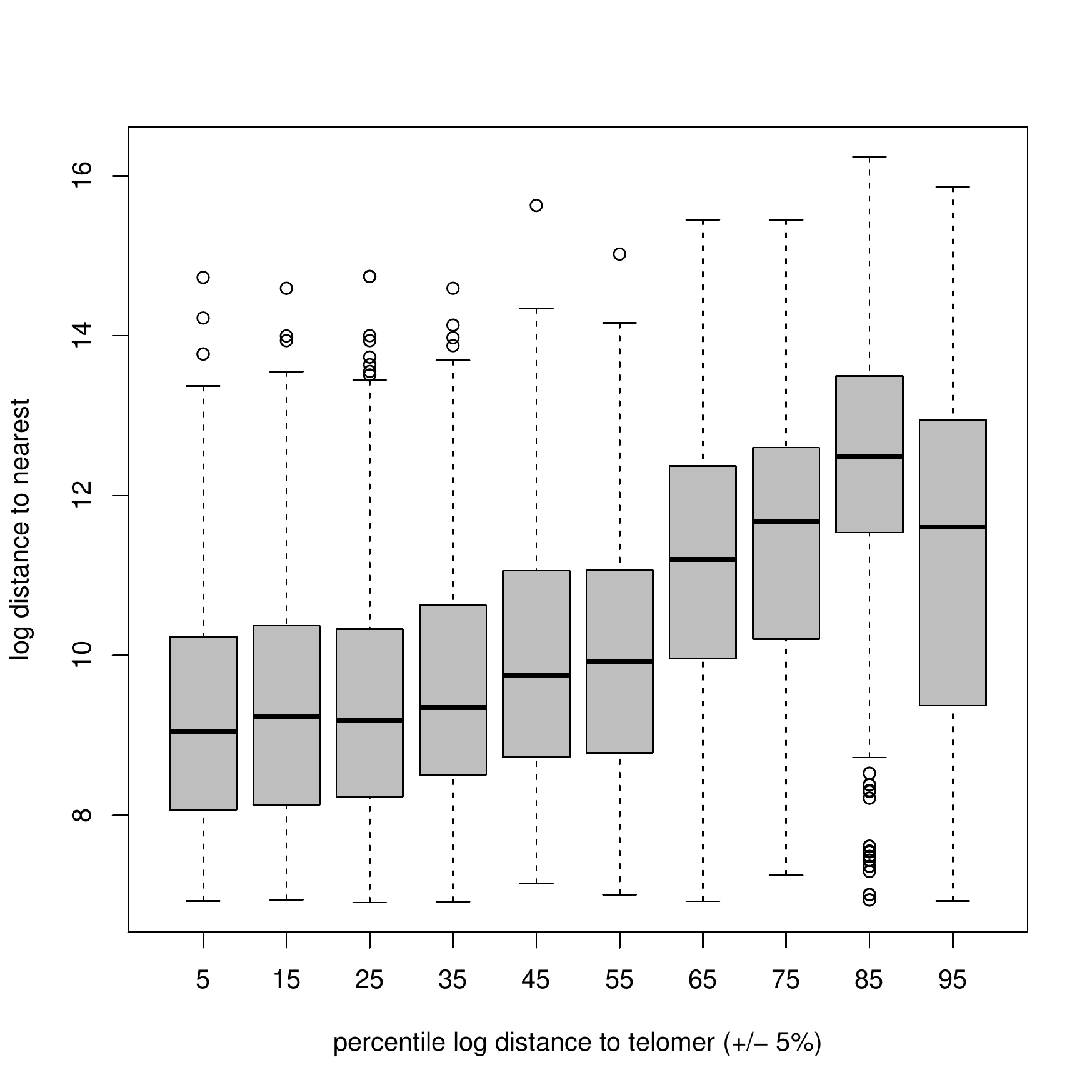}
\end{minipage}
\begin{minipage}{.48\textwidth}
\textsf{D}
\vspace{-.4ex}

\includegraphics[width=.9\textwidth, trim=.1cm .5cm 1cm 1.2cm, clip=true]{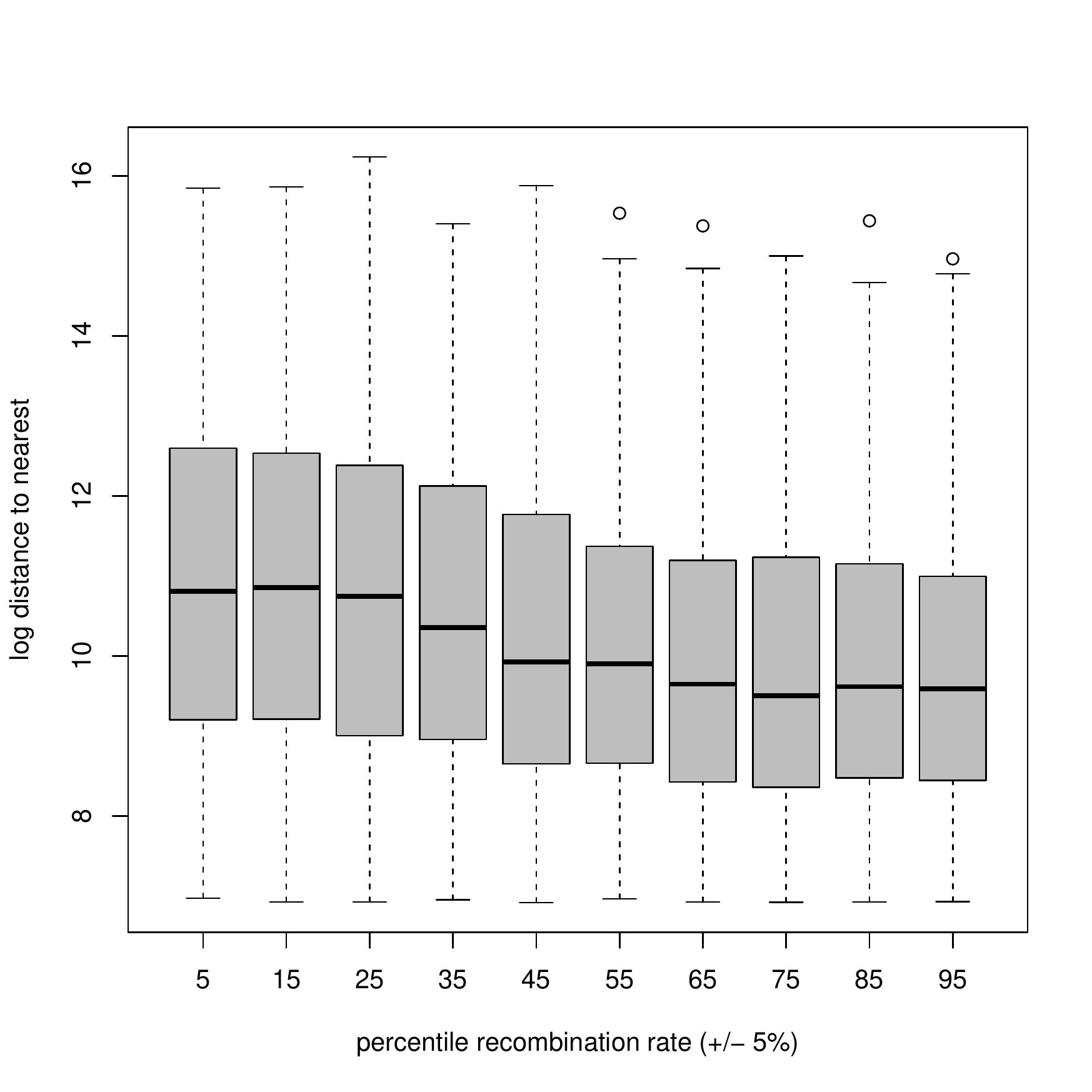}
\end{minipage}
\vspace{1ex}

\caption{
  \label{fig:boxplots} {\bf Distance to telomere and recombination rate
    correlate with gBGC-tract proximity.} This figure shows box plots of
  the distribution of the log distance to nearest in the gBGC tract,
  stratified by log distance to telomere (first column) and 
  recombination rate (second column) for both human (first row) and chimp
  (second row). For both species we observe that gBGC tracts are closer
  together towards the end of chromosomes (panels A and C), and that
  they are further apart in areas of low recombination rate (panels B and
  D). These emprical observations agree with the results of our linear
  modeling analysis (Supplementary Methods).  }

\end{figure}

\clearpage{}
\section*{Supplementary Tables}

\begin{table}[ht]
\begin{center}
\begin{minipage}{.99\textwidth}
\begin{center}
\label{tab:tract_coverage}
\vspace{1ex}
\begin{tabular}{lccccc}
  \hline
  & \bf{\emph{B}=2} & \bf{\emph{B}=3} & \bf{\emph{B}=4} & \bf{\emph{B}=5} & \bf{\emph{B}=10} \\ 
  \hline
  \bf{\emph{B}=2} & 1 & 0.97 & 0.95 & 0.93 & 0.88 \\
  \bf{\emph{B}=3} & 0.29 & 1 & 0.98 & 0.97 & 0.92 \\
  \bf{\emph{B}=4} & 0.19 & 0.64 & 1 & 0.99 & 0.94 \\
  \bf{\emph{B}=5} & 0.13 & 0.46 & 0.72 & 1 & 0.95 \\
  \bf{\emph{B}=10} & 0.06 & 0.20 & 0.31 & 0.44 & 1 \\
  \hline
\end{tabular}
\caption{\textbf{Relative Coverage of Human gBGC Tracts for Various Values of \textbf{\emph{B}}.} Each value in the table represents the fraction of nucleotides in the human gBGC tract predictions for the value of $B$ indicated for the row that also fall in the predictions for the value of $B$ indicated for the column.  The numbers on the main diagonal are one by definition.  The numbers above the main diagonal indicate the coverage of smaller sets (higher $B$) by larger sets (lower $B$), while the numbers below the main diagonal indicate the coverage of larger sets by smaller sets.} 
\end{center}
\end{minipage}
\end{center}
\end{table}

\vspace{1in}
\begin{table}[ht]
\begin{minipage}{.99\textwidth}
\centering
\begin{tabular}{llrcl}
\hline
species&coefficient&value&std. error&$p$-value\\
\hline\noalign{\smallskip}
human & $\beta_1$   & 0.431 & 0.190   &  $<\,10^{-15}$ \\
human & $\beta_2$   & -0.006 & 0.002  & 0.006\\
chimp & $\beta_1$   & 0.613 & 0.141   &  $<\,10^{-15}$ \\
chimp & $\beta_2$   & -0.022 & 0.007  & 9.8$\cdot 10^{-4}$\\
\hline
\end{tabular}
\end{minipage}
\begin{minipage}{.99\textwidth}
\caption{
\label{tab:linmod}
{\bf Distance to telomere and recombination rate are significant factors in
  predicting proximity of gBGC tracts.} This table shows the coefficients
for a linear model in which the log distance-to-nearest gBGC tract
was regressed on the log distance-to-telomere ($\beta_1$) and recombination
rate ($\beta_2$). $p$-values are based on the $T$-test. The
intercept is significant in both species. 
}
\end{minipage}

\end{table}

\end{document}